\documentclass[english,showpacs,prd,aps,superscriptaddress,preprintnumbers,notitlepage]{revtex4}
\usepackage[T1]{fontenc}
\usepackage[latin9]{inputenc}
\usepackage{babel}
\usepackage{amsmath}
\usepackage{amssymb}
\usepackage{graphicx}
\usepackage[unicode=true,
 bookmarks=false,
 breaklinks=false,pdfborder={0 0 1},backref=false,colorlinks=false]
 {hyperref}

\makeatletter

\providecommand{\tabularnewline}{\\}

\usepackage{babel}
\PassOptionsToPackage{normalem}{ulem}
\usepackage{ulem}

\providecommand{\tabularnewline}{\\}

\usepackage{epsfig}
\usepackage{amsfonts}

\makeatother

\begin{document}

\title{Multiplicity dependence of $\chi_{c}$ and $\chi_{b}$ meson production}

\author{Marat Siddikov, Iván Schmidt}

\affiliation{Departamento de Física, Universidad Técnica Federico Santa María,~~~\\
 y Centro Científico - Tecnológico de Valparaíso, Casilla 110-V, Valparaíso,
Chile}
\begin{abstract}
In this paper we analyze in detail the production of the $\chi_{c}$
and $\chi_{b}$ mesons in $pp$ collisions.  Using the color dipole
framework, we estimated the cross-sections in the kinematics of ongoing
and forthcoming experiments, and found that our estimates are in reasonable
agreement with currently available experimental data. We also analyzed
the dependence on multiplicity of co-produced hadrons and found that
it is significantly milder than that of $S$-wave quarkonia. We expect
that the experimental confirmation of this result could constitute an important
test of our understanding of multiplicity enhancement mechanisms in the
production of different quarkonia states.
\end{abstract}

\date{\today}


\maketitle

\section{Introduction}

The standard approach to the description of mesons containing heavy
quarks is based on pomeron-pomeron fusion. The heavy quarks formed
at the partonic level might hadronize into open heavy flavor mesons (e.g.
$D$- or $B$-mesons) or form quarkonia states at later stages of
the collision~\cite{Maciula:2013wg,Chang:1979nn,Baier:1981uk,Berger:1980ni}.
This framework provides reasonable estimates for the total and differential
cross-sections, although it includes some uncertainties due to a rather limited knowledge of the fragmentation
functions of open-flavor mesons or the Long Distance Matrix Elements (LDMEs)
of quarkonia states~\cite{Bodwin:1994jh,Maltoni:1997pt,Brambilla:2008zg,Feng:2015cba,Brambilla:2010cs,Baranov:2015laa,Baranov:2016clx}.
Moreover, this approach recently encountered difficulties with the description
of recent experimental data on multiplicity dependence of co-produced
charged hadrons~\cite{Adam:2015ota,Trzeciak:2015fgz,Ma:2016djk,PSIMULT,Khatun:2019slm,Alice:2012Mult}.
In fact, it was discovered by both the STAR and ALICE collaborations that
the relative yields of the $1S$ quarkonia grow vigorously as function
of multiplicity. This enhancement is seen in $AA$, $pA$~\cite{ALICE:pAStrangeness,ALICE:pAStrangeness2}
and even in $pp$ collisions~\cite{ALICE:2017jyt,Thakur:2018dmp},
which clearly signals that it is not related to collective effects.
Similar enhancement was observed for $D$- and $B$-meson production~\cite{Adam:2015ota}.
As was mentioned in~\cite{Fischer:2016zzs}, these new findings cannot
be easily accommodated in the framework of models based on the two-pomeron
fusion picture and thus potentially could require introduction of
new mechanisms both for $AA$ and $pp$ collisions. 

Recently, in~~\cite{LESI,KMRS,MOSA}, it has been suggested that the
experimentally found multiplicity dependence
could indicate a sizable contribution from multipomeron mechanisms
to charmonia production, and has been shown that the inclusion of 
the three-pomeron mechanism helps to describe the available data.
Also, it was shown in~\cite{Schmidt:2020fgn} that in the $D$-meson
production case the three-pomeron mechanism gives a sizable contribution,
which might be responsible for up to 40 percent of all the produced $D$-mesons.
The inclusion of this mechanism improves the overall description of the
data, as well as allows to describe its multiplicity dependence. In
order to understand better the role of the three-pomeron fusion mechanism
in heavy charm production, it has been suggested in~\cite{Siddikov:2020pjh}
to study the multiplicity dependence of diffractive production.
Although the predicted cross-section is smaller than for the inclusive
case, we expect that its multiplicity dependence could be studied
during the High Luminosity Run 3 at the LHC (HL-LHC mode)~\cite{ATLAS:2013hta,Apollinari:HLLHC,LaRoccaRiggi}.

In order to understand better the microscopic mechanisms of multiplicity
enhancement in heavy quarkonia production, in this paper we suggest
to study the multiplicity dependence of co-produced hadrons in the production
of $P$-wave quarkonia, \emph{e.g}. the lightest $\chi_{c}$ and $\chi_{b}$-mesons.
The production of $\chi_{c}$ mesons has been extensively
studied in the $k_{T}$ factorization approach in~\cite{Likhoded:2014kfa,Baranov:2015yea,Jia:2014jfa,Hagler:2000dd,Babiarz:2020jkh,Cisek:2017gno,Diakonov:2012vb,Boer:2012bt},
where it was found that a two-pomeron (two-gluon) fusion mechanism provides
a good description of the available data on its rapidity and transverse
momentum dependence. Moreover, the color octet mechanism for $P$-wave quarkonia
gives a small contribution due to smallness of the Long Distance matrix
Elements (LDMEs)~\cite{Likhoded:2014kfa,Baranov:2015yea,Jia:2014jfa},
so this fact minimizes the inherent uncertainty related to this mechanism.
On the other hand, due to spin-orbital interactions, the $P$-wave quarkonia show up
as a triplet of states with different angular momenta $J=0,\,1$,
or 2. Independent studies of experimental cross-sections of each of
these states provides a sensitive test of the underlying production
mechanism. For this reason, we believe that $P$-wave quarkonia are
ideally suited for the study of the multiplicity enhancement mechanisms.
As we will show below, in the high energy limit the three-pomeron
mechanism does not contribute to $P$-wave quarkonia production, so
we expect that the multiplicity dependence for $\chi_{c}$ and $\chi_{b}$
mesons should be significantly milder than that of $S$-wave quarkonia.
Since we are interested in the multiplicity dependence, instead of $k_{T}$-factorization
we will use a color dipole framework (also known as CGC/Saturation
or CGC/Sat)~\cite{GLR,McLerran:1993ni,McLerran:1993ka,McLerran:1994vd,MUQI,MV,gbw01:1,Kopeliovich:2002yv,Kopeliovich:2001ee}.
The generalization of this framework to high-multiplicity events is
well-known from the literature~\cite{KOLEB,KLN,DKLN,Kharzeev:2000ph,Kovchegov:2000hz,LERE,Lappi:2011gu,Ma:2018bax}.

The paper is structured as follows. In the next section~\ref{sec:ProdMec}
we describe a framework for $\chi_{cJ}$ and $\chi_{bJ}$ quarkonia
production. In Section~\ref{sec:Numer} we make numerical estimates
of the cross-sections, compare them with available experimental data
and make predictions for future experiments. In Section~\ref{sec:Multiplicity}
we evaluate the dependence on multiplicity. Finally, in Section~\ref{sec:Concusion}
we draw conclusions.

\section{Production mechanisms of $P$-wave quarkonia}

\label{sec:ProdMec}

\begin{figure}
\includegraphics[width=9cm]{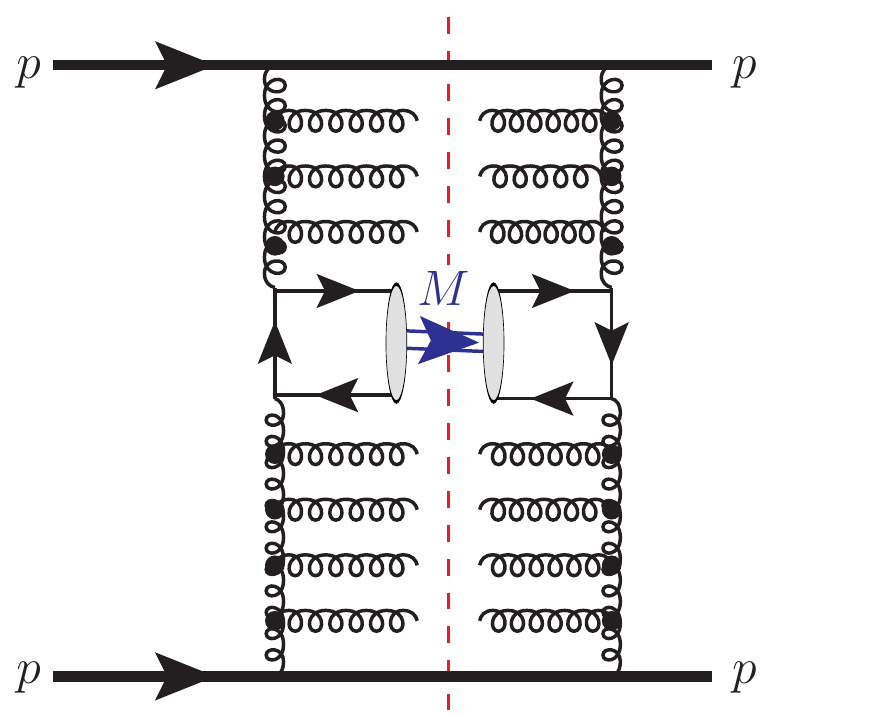}\includegraphics[width=9cm]{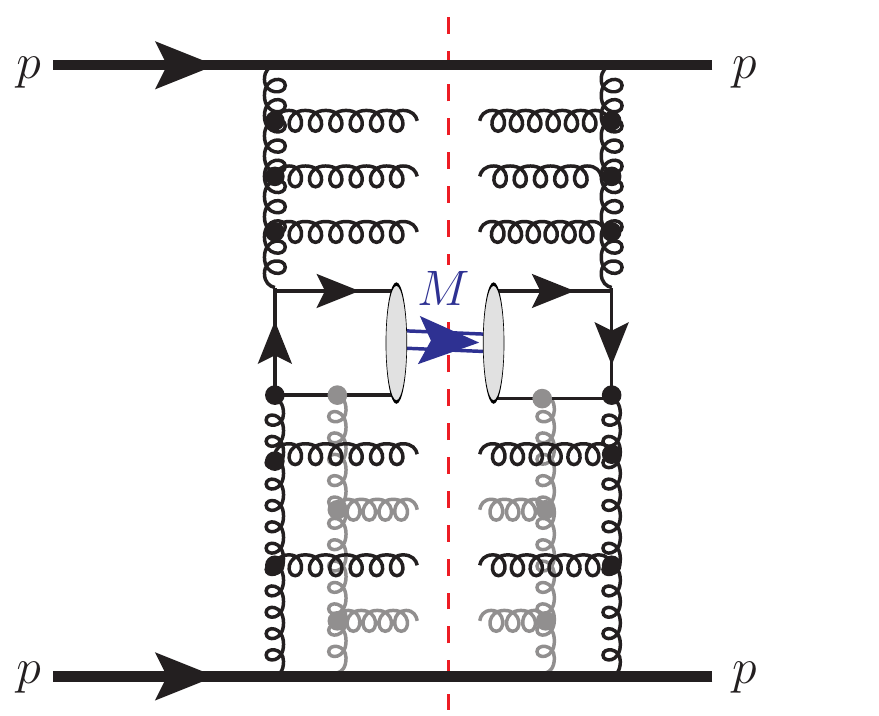}\caption{\label{fig:DipoleCrossSections-2Pom}Left plot: The leading order
contribution to the cross-section of $P$-wave meson production
via the two-pomeron fusion mechanism. The diagram includes two cut pomerons
(upper and lower gluon ladders). Right plot: Possible contributions
of the 3-reggeon mechanism (additional reggeon shown with gray color).
As explained in the text, if the two lower gluons reggeize indepedendently
(form two pomerons), then such contribution for $P$-wave quarkonia
would vanish. In both plots the vertical dashed line stands for unitarity
cuts. The produced meson $M$ is shown with a double line and arrow.
A summation over all possible permutations of gluon vertices in the
heavy quark line/loop is implied.}
\end{figure}
In high energy kinematics, the inclusive production gets its dominant
contribution from the fusion of two pomerons, which for heavy quarkonia
production is given by the diagram shown in the left panel of
Figure~\ref{fig:DipoleCrossSections-2Pom}. In the rest frame of
one of the protons this process might be viewed as a fluctuation of
the incoming virtual gluon into a heavy $\bar{Q}Q$ pair, with subsequent
scattering of the $\bar{Q}Q$ dipole with the target proton. In the
kinematics of LHC experiments the average light-cone momentum fractions
$x_{1,2}$ carried by gluons are very small ($\ll1$), and the gluon
densities are enhanced. This enhancement implies
that there could be sizable corrections from multiple pomeron exchanges
between the heavy dipole and the target, which are formally suppressed
in the heavy quark mass limit. For this reason, instead of a hard process
on individual partons it is more appropriate to use the color dipole
framework (also known as CGC/Sat)~\cite{GLR,McLerran:1993ni,McLerran:1993ka,McLerran:1994vd,MUQI,MV,gbw01:1,Kopeliovich:2002yv,Kopeliovich:2001ee}.
The color dipoles are eigenstates of interaction at high energies,
and for this reason can be used as universal elementary building
blocks automatically accumulating both the hard and soft fluctuations~\cite{Nikolaev:1994kk}.
In fact, the light-cone color dipole framework has been successfully applied
to phenomenological descriptions of both hadron-hadron and lepton-hadron
collisions~\cite{Kovchegov:1999yj,Kovchegov:2006vj,Balitsky:2008zza,Kovchegov:2012mbw,Balitsky:2001re,Cougoulic:2019aja,Aidala:2020mzt,Ma:2014mri}.
Another advantage of the CGC/Sat (color dipole) framework is that
it allows a relatively straightforward extension for the description
of high-multiplicity events, as discussed in~\cite{KOLEB,KLN,DKLN,Kharzeev:2000ph,Kovchegov:2000hz,LERE,Lappi:2011gu,Ma:2018bax}.
The cross-section of quarkonia production process, shown in Figure~\ref{fig:DipoleCrossSections-2Pom},
in the dipole approach is given by (see details in Appendix~\ref{sec:Derivation})
\begin{eqnarray}
\frac{d\sigma_{M}\left(y,\,\sqrt{s}\right)}{dy\,d^{2}p_{T}} & = & \,\int d^{2}k_{T}x_{1}\,g\left(x_{1},\,\boldsymbol{p}_{T}-\boldsymbol{k}_{T}\right)\int_{0}^{1}dz_{1}\int_{0}^{1}dz_{2}\int\frac{d^{2}r_{1}}{4\pi}\,\int\frac{d^{2}r_{2}}{4\pi}\int d^{2}\boldsymbol{b}_{21}e^{i\boldsymbol{b}_{21}\cdot\boldsymbol{k}_{T}}\times\label{FD1}\\
 & \times & \left\langle \Psi_{\bar{Q}Q}^{\dagger}\left(r_{1},\,z_{1}\right)\,\Psi_{M}\left(r_{1},\,z_{1}\right)\right\rangle \left\langle \Psi_{\bar{Q}Q}^{\dagger}\left(r_{2},\,z_{2}\right)\,\Psi_{M}\left(r_{2},\,z_{2}\right)\right\rangle ^{*}N_{M}\left(x_{2};z_{1,}\,\boldsymbol{r}_{1};\,z_{2},\,\boldsymbol{r}_{2};\,\boldsymbol{b}_{21}\right)\nonumber \\
 & + & \left(x_{1}\leftrightarrow x_{2}\right),\nonumber 
\end{eqnarray}
\begin{eqnarray}
N_{M}\left(x;\,z_{1,}\,\boldsymbol{r}_{1};\,z_{2},\,\boldsymbol{r}_{2};\,\boldsymbol{b}_{21}\right) & = & N\left(x,\,\boldsymbol{b}_{21}+\bar{z}_{2}\boldsymbol{r}_{2}+z_{1}\boldsymbol{r}_{1}\right)+N\left(x\,\boldsymbol{b}_{21}-\bar{z}_{1}\boldsymbol{r}_{1}-z_{2}\boldsymbol{r}_{2}\right)-\label{eq:N_MDef}\\
 &  & -N\left(x,\,\boldsymbol{b}_{21}+\bar{z}_{2}\boldsymbol{r}_{2}-\bar{z}_{1}\boldsymbol{r}_{1}\right)-N\left(x,\,\boldsymbol{b}_{21}-\bar{z}_{1}\boldsymbol{r}_{1}-z_{2}\boldsymbol{r}_{2}\right)\nonumber \\
x_{1,2} & \approx & \frac{\sqrt{m_{M}^{2}+\langle p_{\perp M}^{2}\rangle}}{\sqrt{s}}e^{\pm y}
\end{eqnarray}
where $y$ and $p_{T}$ are the rapidity and transverse momenta of the
produced quarkonia in the center-of-mass frame of the colliding protons;
$\left(z_{i},\,\boldsymbol{r}_{i}\right)$ are the light-cone fractions
of the quark and the transverse separation between quarks inside the dipole
(with subindices $i=1,2$, standing for amplitude and its complex conjugate
respectively); $\boldsymbol{b}_{21}$ is the difference of impact
parameters of the dipoles in the amplitude and its conjugate. We also
use the notation $\Psi_{M}(r,\,z)$ for the light-cone wave function
of quarkonium $M$ ($M=\chi_{c},\,\chi_{b}$), and $\Psi{}_{\bar{Q}Q}$
for the quark-antiquark component of the gluon light-cone wave function
(for the sake of completeness both are discussed in detail in Appendix~\ref{sec:WFs}).
The amplitude $N_{M}$ depends on a linear combination of forward
dipole scattering amplitudes $N\left(y,\,\boldsymbol{r}\right)\equiv\int d^{2}\boldsymbol{b}\,N\left(y,\,\boldsymbol{r},\,\boldsymbol{b}\right)$.
The notation $\,x_{g}g\left(x_{g},\,\boldsymbol{k}_{T}\right)$ is
used for the unintegrated gluon PDF. The expression for the $p_{T}$-integrated cross-section
has a similar structure and differs only by the replacement
of the gluon uPDF $x_{1}\,g\left(x_{1},\,\boldsymbol{p}_{T}-\boldsymbol{k}_{T}\right)$
by the integrated PDF $x_{1}g\left(x_{1},\mu_{F}\right)$, taken
at the scale $\mu_{F}\,\approx2\,m_{Q}$.  The integrated gluon PDF
$x_{1}g\left(x_{1},\,\mu_{F}\right)$ in the CGC/Sat approach is closely
related to the dipole scattering amplitude $N\left(y,\,\boldsymbol{r}\right)$
introduced earlier as~\cite{KOLEB,THOR}
\begin{equation}
\frac{C_{F}}{2\pi^{2}\bar{\alpha}_{S}}N\left(y,\,\boldsymbol{r}\right)=\int\frac{d^{2}k_{T}}{k_{T}^{4}}\phi\left(y,k_{T}\right)\,\Bigg(1-e^{i\boldsymbol{k}_{T}\cdot\boldsymbol{r}}\Bigg);~~~~x\,g\left(x,\,\mu_{F}\right)=\int_{0}^{\mu_{F}}\frac{d^{2}k_{T}}{k_{T}^{2}}\phi\left(x,\,k_{T}\right),\label{GN1}
\end{equation}
where $y=\ln(1/x)$. Eq. (\ref{GN1}) might be inverted and gives
the gluon PDF in terms of the dipole amplitude, 
\begin{equation}
x\,g\left(x,\,\mu_{F}\right)\,\,=\,\,\frac{C_{F}\mu_{F}}{2\pi^{2}\bar{\alpha}_{S}}\int d^{2}r\,\frac{J_{1}\left(r\,\mu_{F}\right)}{r}\nabla_{r}^{2}N\left(y,\,\boldsymbol{r}\right).\label{GN2}
\end{equation}
This result allows us to rewrite~(\ref{FD1}) entirely
in terms of the dipole amplitude $N$. 

\begin{figure}
\includegraphics[width=9cm]{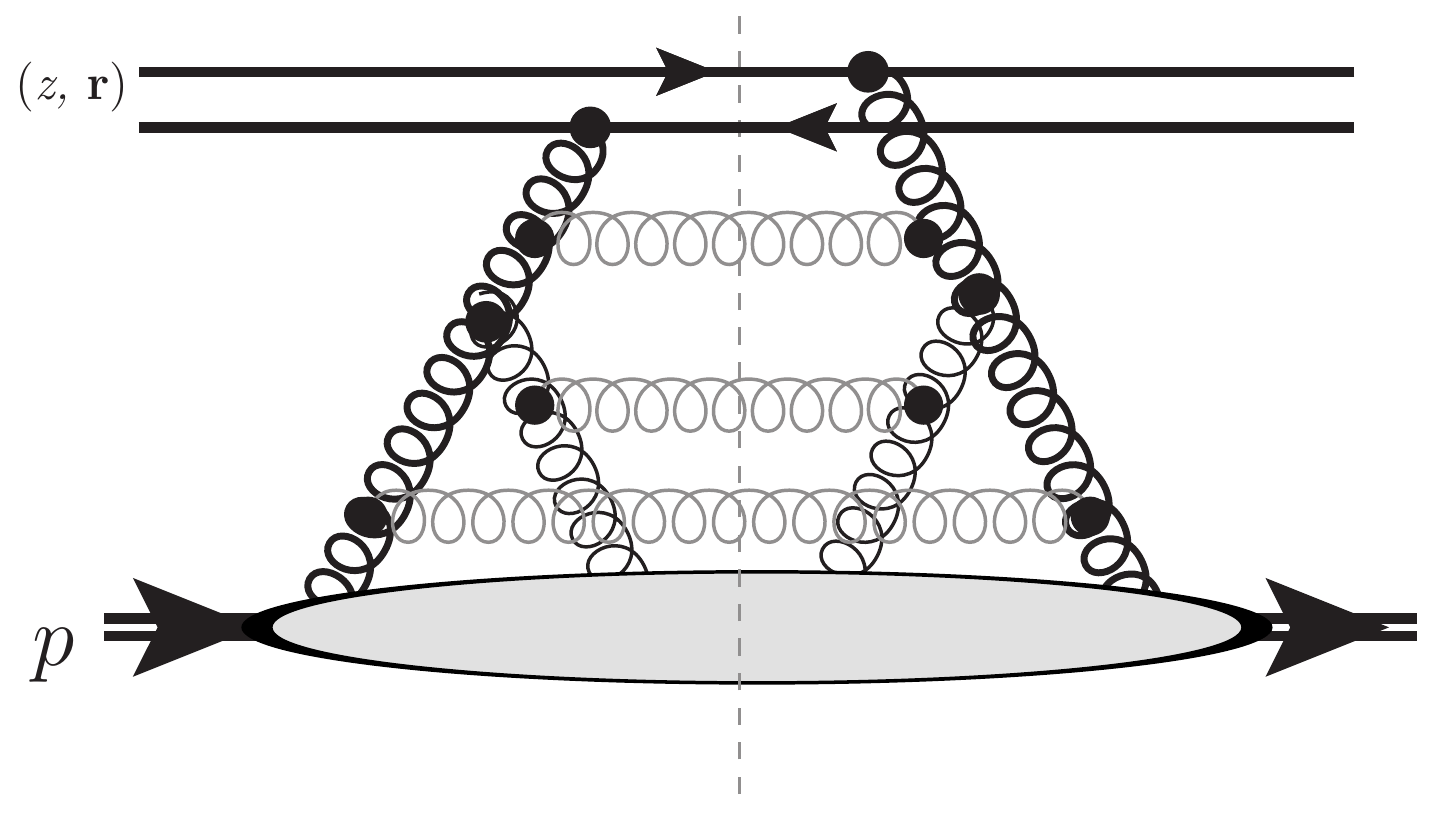}\includegraphics[width=9cm]{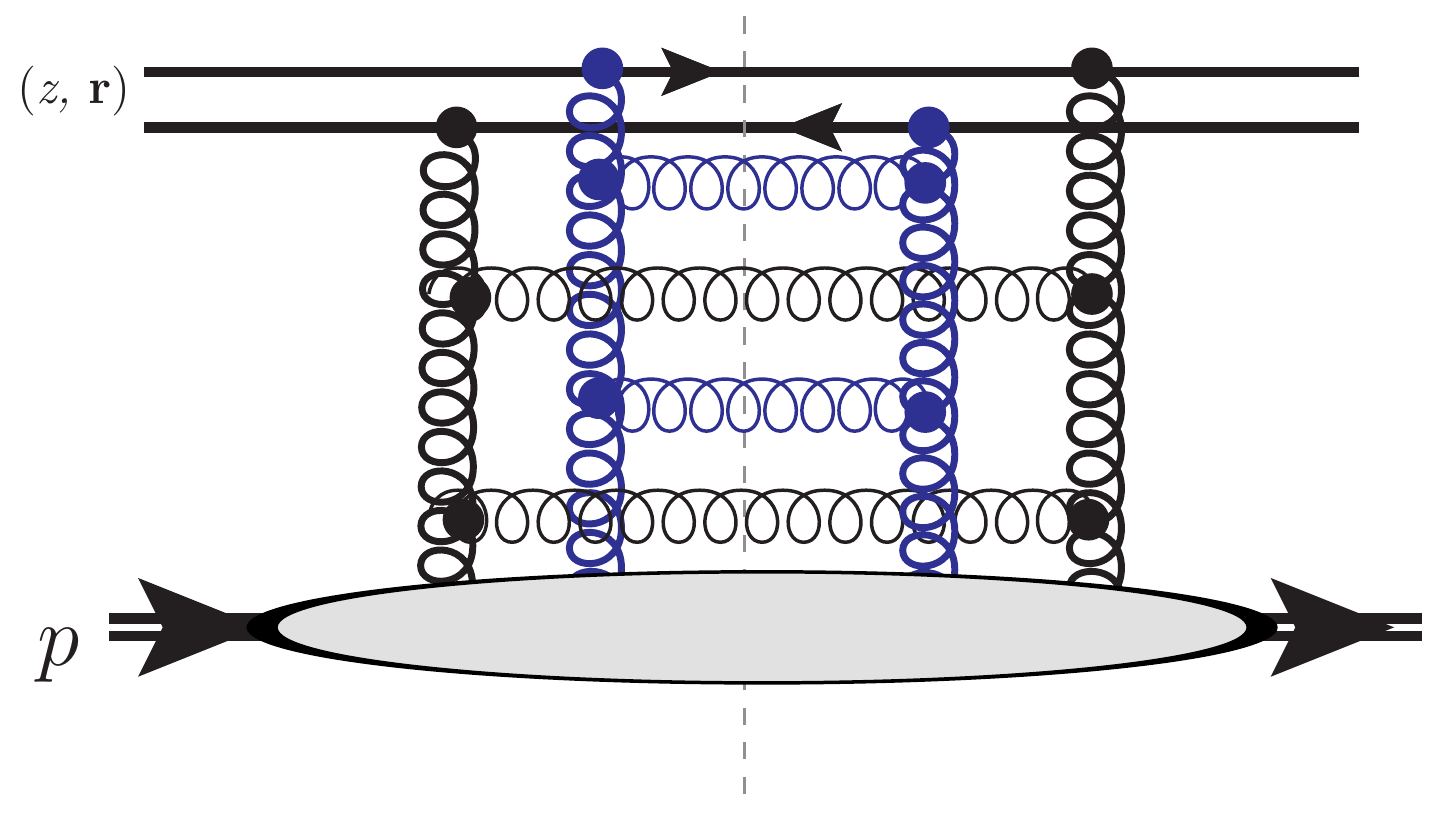}

\caption{\label{fig:DipoleCrossSections}Left plot: A typical fan diagram taken
into account in the CGC parametrization~~\cite{Iancu:2003ge,RESH,Kowalski:2006hc,Watt:2007nr}
of the \emph{color singlet} dipole cross-section $N(z,\,r)$ (resummation
of all possible tree-like topologies is implied).~ Right plot: The
BFKL ladder diagrams resummed in the IP-Sat (\emph{b}-Sat) parametrization~\cite{Kowalski:2003hm,Rezaeian:2012ji}.
In both plots a vertical dashed grey line stands for the unitarity
cut, and the blob in the lower part is the hadronic target (proton); two
fermionic lines in the upper part of the blob stand for the dipole
of transverse size $\boldsymbol{r}$. }
\end{figure}

Now we would like to discuss the contributions of the three-pomeron
mechanisms shown in the right panel of Figure~\ref{fig:DipoleCrossSections-2Pom}.
Usually it is expected that such contributions are suppressed in the heavy
quark mass limit by $\alpha_{s}\left(m_{Q}\right)$, and in certain
cases additionally by $\Lambda_{{\rm QCD}}^{2}/m_{Q}^{2}$. However,
for charmonia the latter parameter might not be very small, and for
this reason some corrections could be sizable. As was demonstrated
in~\cite{Siddikov:2019xvf,Levin:2018qxa}, this indeed happens in the
case of $1S$ charmonia production, and these contributions are especially
important in the events with large multiplicity of co-produced hadrons.
For this reason we also need to estimate them in the case of $\chi_{c}$
and $\chi_{b}$ production. In the color dipole framework it is usually
assumed that a universal dipole cross-section takes into account all
such contributions. However, in phenomenological parametrizations
usually such contributions are either taken into account using additional
simplifying assumptions or disregarded altogether. 
For example, a widely used phenomenological parametrization
``CGC'', suggested in~\cite{Iancu:2003ge,RESH,Kowalski:2006hc,Watt:2007nr},
was inspired by a solution of the Balitsky-Kovchegov (BK) equation
and effectively resumms only fan diagrams shown in the left panel of
Figure~\ref{fig:DipoleCrossSections}. This parametrization does
not take into account the three-pomeron contributions at all. The
alternative IP-Sat parametrization~\cite{Kowalski:2003hm,Rezaeian:2012ji},
inspired by a Glauber-like approach, resumms the set of BFKL
ladder diagrams shown in the right panel of Figure~\ref{fig:DipoleCrossSections}.
A central assumption of this approach is that the interaction of each
BFKL ladder (pomeron) with a dipole of size $r$ is given by $\sim\alpha_{s}\left(\mu^{2}\right)r^{2}x\,g(x)$,
which might work for small color-singlet dipoles, but in the general case
it requires more careful treatment. For this reason, in general we
cannot extract the contribution of the three-pomeron mechanism by just
expanding the dipole amplitude, and instead we should evaluate it explicitly. 

As was demonstrated in~\cite{LESI}, the contribution due the to the three-pomeron
mechanism has a structure similar to~(\ref{FD1}), although we need
to replace the amplitude $N_{M}$ defined in~(\ref{eq:N_MDef}) with
a new dipole amplitude $\tilde{N}_{M}$, which explicitly takes into
account two-Pomeron interaction of the dipole with the target proton.
As was shown in~\cite{Korchemsky:2001nx}, at high energies the dominant
configuration (largest intercept) comes from the configuration when
each pomeron reggeizes independently, which implies that the amplitude
$\tilde{N}_{M}$ has the form 
\begin{equation}
\tilde{N}_{M}\left(x_{2};z_{1,}\,\boldsymbol{r}_{1};\,z_{2},\,\boldsymbol{r}_{2};\,\boldsymbol{b}_{21}\right)=\kappa N_{M}^{2}\left(x_{2};z_{1,}\,\boldsymbol{r}_{1};\,z_{2},\,\boldsymbol{r}_{2};\,\boldsymbol{b}_{21}\right)\label{eq:N3Def}
\end{equation}
where $\kappa$ is a numerical prefactor, which depends on the transverse
profile (impact parameter dependence) of the dipole amplitude $N$,
\begin{equation}
\kappa=\int d^{2}\boldsymbol{b}\,T^{2}(\boldsymbol{b}),\quad T(\boldsymbol{b})\approx\lim_{r\to0}\frac{N(x,\,\boldsymbol{r},\,\boldsymbol{b})}{\int d^{2}b\,N(x,\,\boldsymbol{r},\,\boldsymbol{b})}.
\end{equation}
From the structure of~(\ref{FD1}) and the symmetry properties of
the wave function of the $P$-wave quarkonia, we can see that such
contribution vanishes. The configurations which might give nonzero
contributions require antisymmetrization over color indices of both
pomerons, thus forming a multireggeon state. It has been shown in the
literature that such configurations have smaller intercepts and are
suppressed at high energies~\cite{Korchemsky:2001nx}. For this reason
we may consider that in LHC kinematics there is no three-pomeron contributions
like those shown in the right panel of the Figure~\ref{fig:DipoleCrossSections-2Pom}.

Finally, we will discuss briefly possible contributions of the color
octet mechanism~\cite{Bodwin:1994jh,Maltoni:1997pt,Brambilla:2008zg,Feng:2015cba,Brambilla:2010cs,Baranov:2015laa,Baranov:2016clx},
which might be relevant in large-$p_{T}$ kinematics. For the
$P$-wave quarkonia the relevant contribution is controlled by the
Long Distance Matrix Element (LDME) $\mathcal{O}^{\chi_{c}}\left[^{3}S_{1}^{(8)}\right]$.
The analyses available from the literature~\cite{Likhoded:2014kfa,Baranov:2015yea,Jia:2014jfa}
conclude that the value of this LDME is very small, although the estimates
of its exact value vary significantly, between $4.78\times10^{-5}$
and $2.01\times10^{-3}\,{\rm GeV}^{3}$. In view of these findings,
in what follows we will simply omit the contribution of the color
octet mechanism.

\section{Numerical estimates}

\label{sec:Numer}In the CGC/saturation approach, the dipole amplitude
$N\left(y,\,\vec{r},\,\vec{b}\right)$ is expected to satisfy the
non-linear Balitsky-Kovchegov equation~\cite{BK} for the dipoles
of small size $r$. In the saturation region this solution should
exhibit a geometric scaling, being a function of one variable $\tau\,=\,r^{2}\,Q_{s}^{2}$,
where $Q_{s}$ is the saturation scale~\cite{GS1,GS2,GS3,LETU}.
Such behavior is implemented in different phenomenological parametrizations
available from the literature. One of such parametrizations which
we will use for our numerical estimates is the CGC parametrization~\cite{RESH}, 

\begin{align}
N\left(x,\,\vec{\boldsymbol{r}}\right) & =\sigma_{0}\times\left\{ \begin{array}{cc}
N_{0}\,\left(\frac{r\,Q_{s}(x)}{2}\right)^{2\gamma_{{\rm eff}}(r)}, & r\,\le\frac{2}{Q_{s}(x)}\\
1-\exp\left(-\mathcal{A}\,\ln\left(\mathcal{B}r\,Q_{s}\right)\right), & r\,>\frac{2}{Q_{s}(x)}
\end{array}\right.~,\label{eq:CGCDipoleParametrization}\\
 & \mathcal{A}=-\frac{N_{0}^{2}\gamma_{s}^{2}}{\left(1-N_{0}\right)^{2}\ln\left(1-N_{0}\right)},\quad\mathcal{B}=\frac{1}{2}\left(1-N_{0}\right)^{-\frac{1-N_{0}}{N_{0}\gamma_{s}}},\\
 & Q_{s}(x)=\left(\frac{x_{0}}{x}\right)^{\lambda/2},\,\,\gamma_{{\rm eff}}(r)=\gamma_{s}+\frac{1}{\kappa\lambda Y}\ln\left(\frac{2}{r\,Q_{s}(x)}\right),\label{eq:gamma_eff}\\
 & \gamma_{s}=0.762,\quad\lambda=0.2319,\quad\sigma_{0}=21.85\,{\rm mb},\quad x_{0}=6.2\times10^{-5}\\
 & Y=\ln\left(1/x\right)
\end{align}

We would like to start our discussion of results from a comparison of the
predicted $p_{T}$-dependence of the cross-sections with experimental
data. The cross-section of $\chi_{c}$ production is smaller than
the cross-section of $J/\psi$, for this reason there is much less
experimental data available from the literature. Since $\chi_{cJ}$
is usually detected via the $\chi_{cJ}\to\gamma+J/\psi$ radiative
decay channel, the experimental data are traditionally presented for
the product of the cross-section onto the branching fraction $\mathcal{B}\left(\chi_{cJ}\right)\equiv Br\left(\chi_{cJ}\to\gamma+J/\psi\right)Br\left(J/\psi\to\mu^{+}\mu^{-}\right)$.
For the $\chi_{bJ}$ mesons we use a similar product of branching
fractions with $\Upsilon(1S)$ instead of $J/\psi$. The values of
the branching fractions are known from~\cite{Zyla:2020zbs} and for
the sake of completeness are shown in Table~\ref{tab:Branchings}.
As we can see, the values of $\mathcal{B}\left(\chi_{c0}\right)$
and $\mathcal{B}\left(\chi_{b0}\right)$ are extremely small compared
to the other channels. For this reason observation of these states
via radiative decays into $1S$ quarkonia is very difficult, and all
the available data are given for $\chi_{c1}$ and $\chi_{c2}$ mesons.

\begin{table}

\begin{tabular}{|c|c|c|c|}
\hline 
 & $J=0$ & $J=1$ & $J=2$\tabularnewline
\hline 
$\mathcal{B}\left(\chi_{cJ}\right)$ & 0.08~\% & 2.02~\% & 1.12~\%\tabularnewline
\hline 
$\mathcal{B}\left(\chi_{bJ}\right)$ & 0.05\% & 0.87~\% & 0.44~\%\tabularnewline
\hline 
\end{tabular}\caption{\label{tab:Branchings}Values of the product of branching fractions
$\mathcal{B}\left(\chi_{cJ}\right)\equiv Br\left(\chi_{cJ}\to\gamma+J/\psi\right)Br\left(J/\psi\to\mu^{+}\mu^{-}\right)$
and $\mathcal{B}\left(\chi_{bJ}\right)\equiv Br\left(\chi_{cJ}\to\gamma+\Upsilon(1S)\right)Br\left(\Upsilon(1S)\to\mu^{+}\mu^{-}\right)$,
as given in~\cite{Zyla:2020zbs}.}
\end{table}

\begin{figure}
\includegraphics[width=9cm]{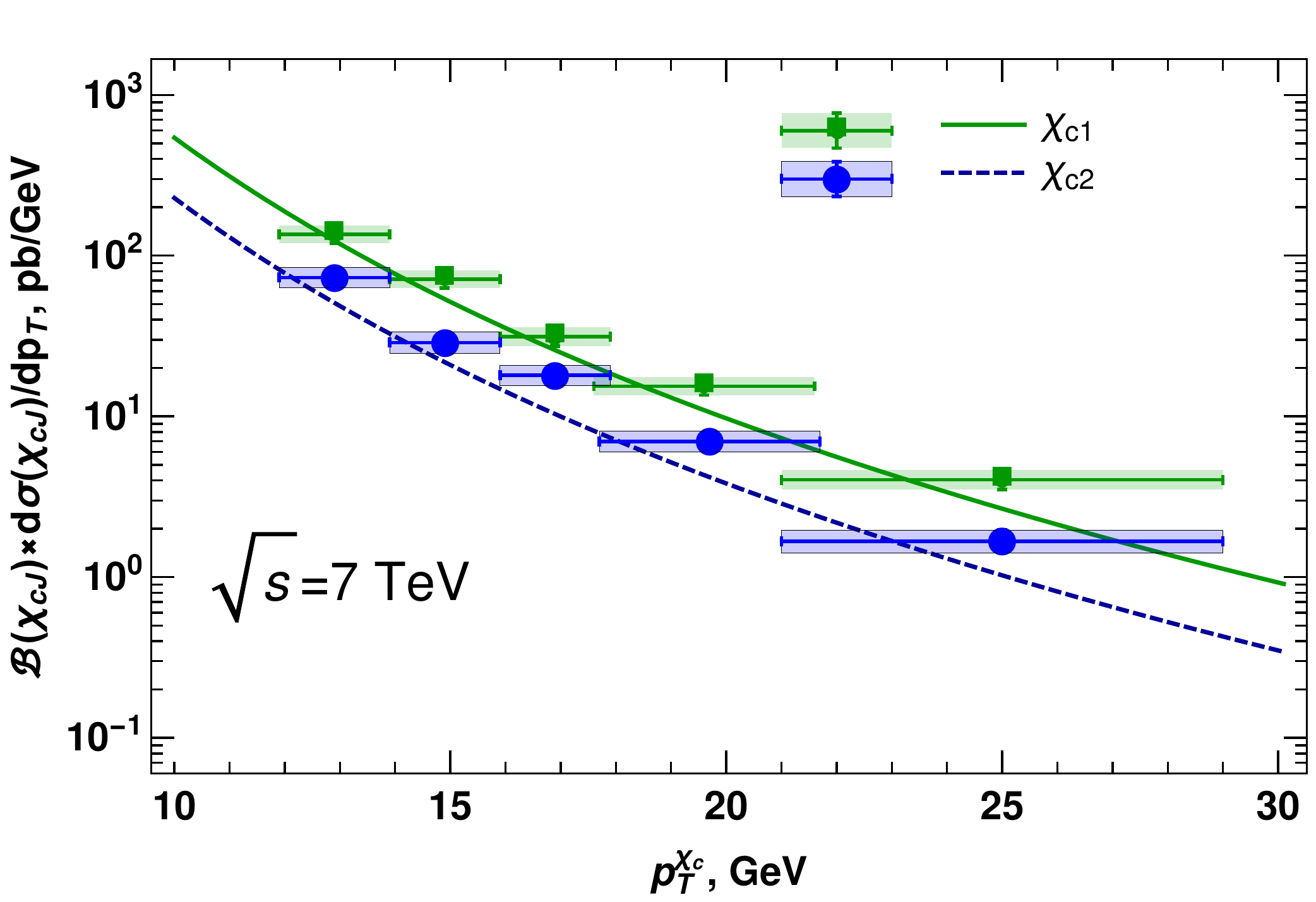}\includegraphics[width=9cm]{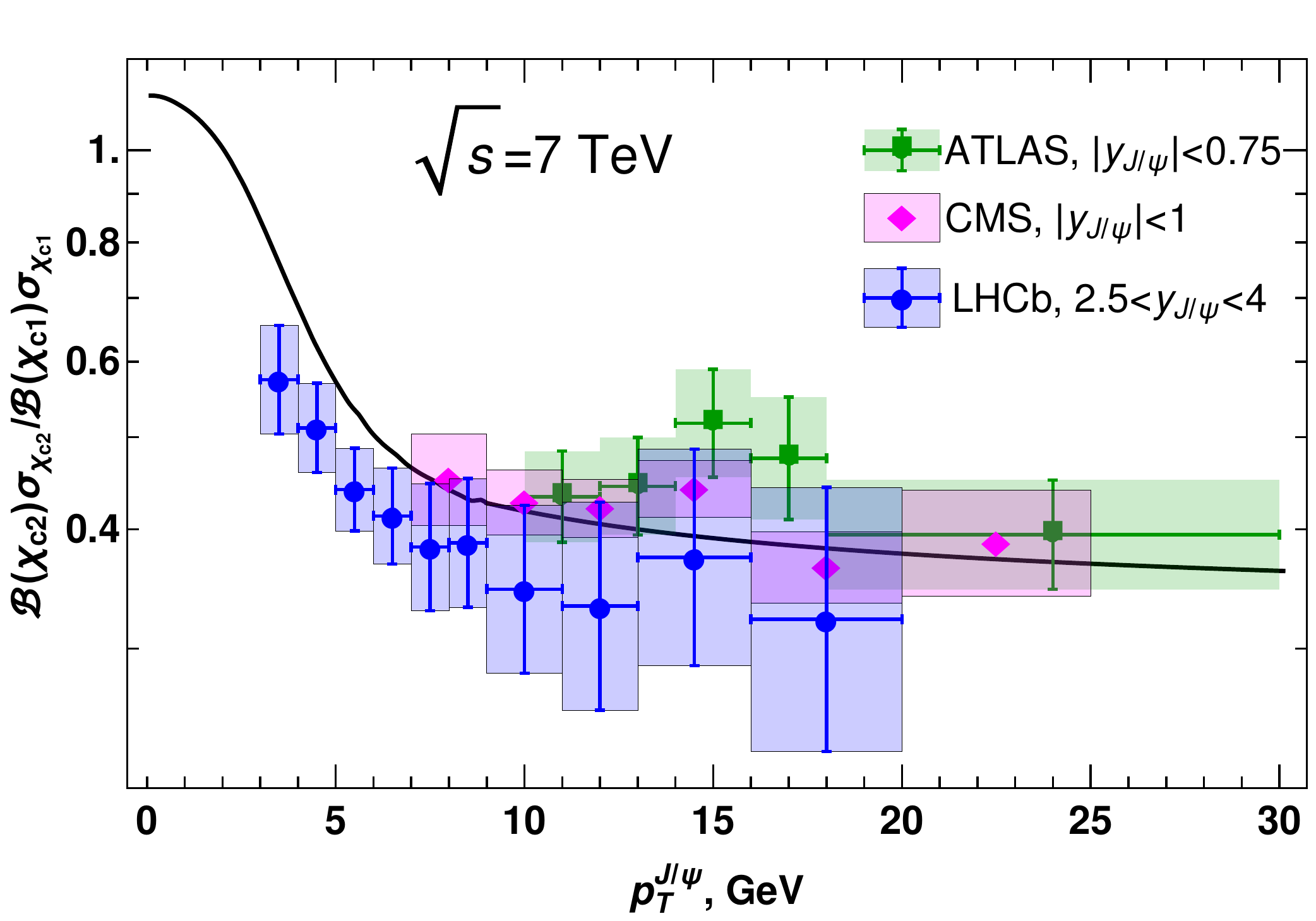}

\caption{\label{fig:ComparisonDataChic}Left: Comparison of the predicted $p_{T}$-dependence
for the $\chi_{c1}$ and $\chi_{c2}$ cross-sections. Experimental
data are from ATLAS~\cite{ATLAS:2014ala}. Right: Comparison of the
model predictions for the ratio of $\chi_{c1}$ and $\chi_{c2}$ cross-sections
at central rapidities, with experimental data from ATLAS~\cite{ATLAS:2014ala},
CMS~\cite{Chatrchyan:2012ub} and LHCb~\cite{Aaij:2013dja} experiments.
We added for comparison the LHCb data measured at off-forward rapidities,
because we expect that the suppression of the cross-sections of $\chi_{c1}$
and $\chi_{c2}$ production at off-forward rapidities will be the
same and thus will cancel in the ratio. For better visibility we use a
logarithmic scale in the vertical axis.}

\label{pT-dependence} 
\end{figure}

\begin{figure}
\includegraphics[width=9cm]{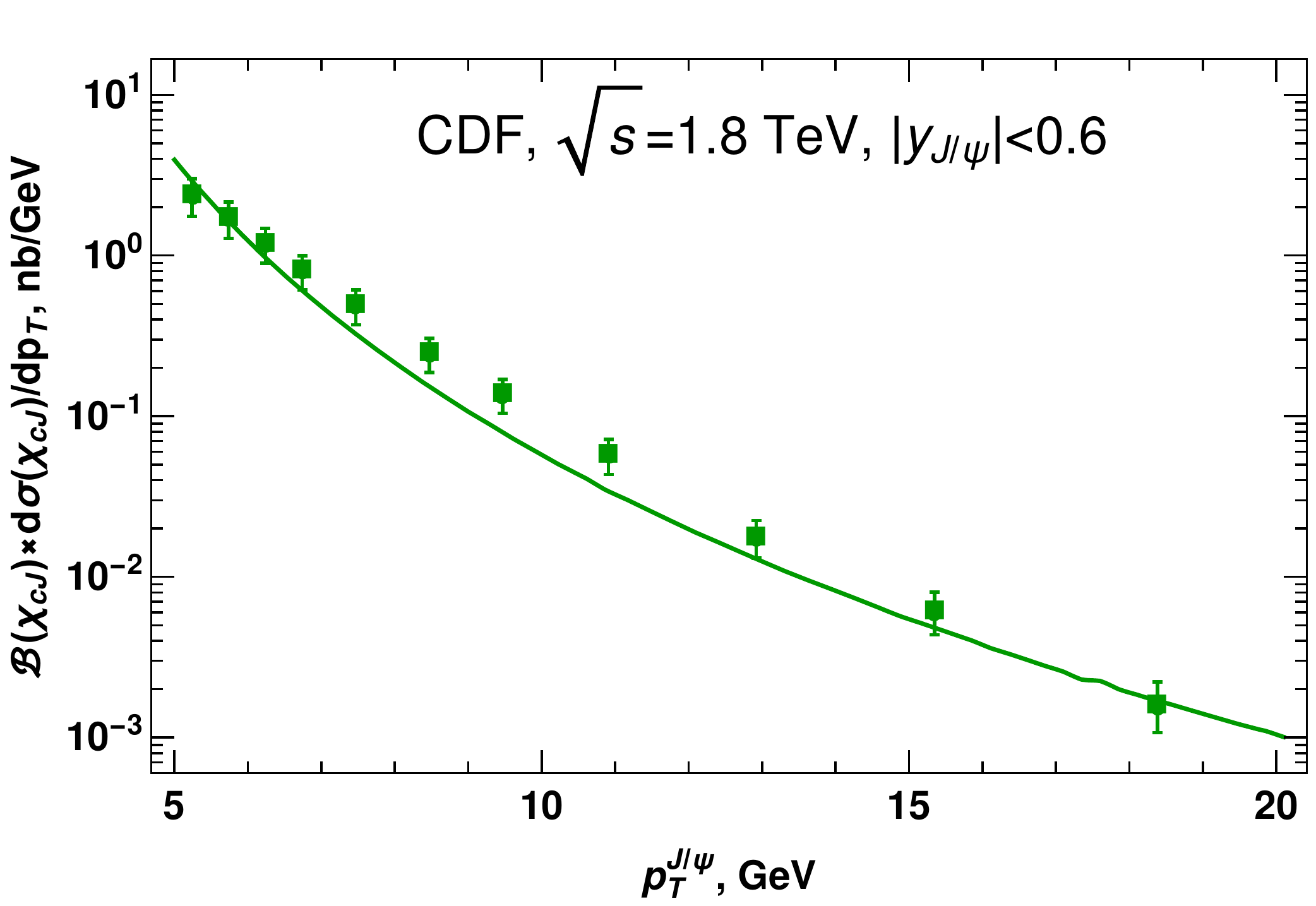}\includegraphics[width=9cm]{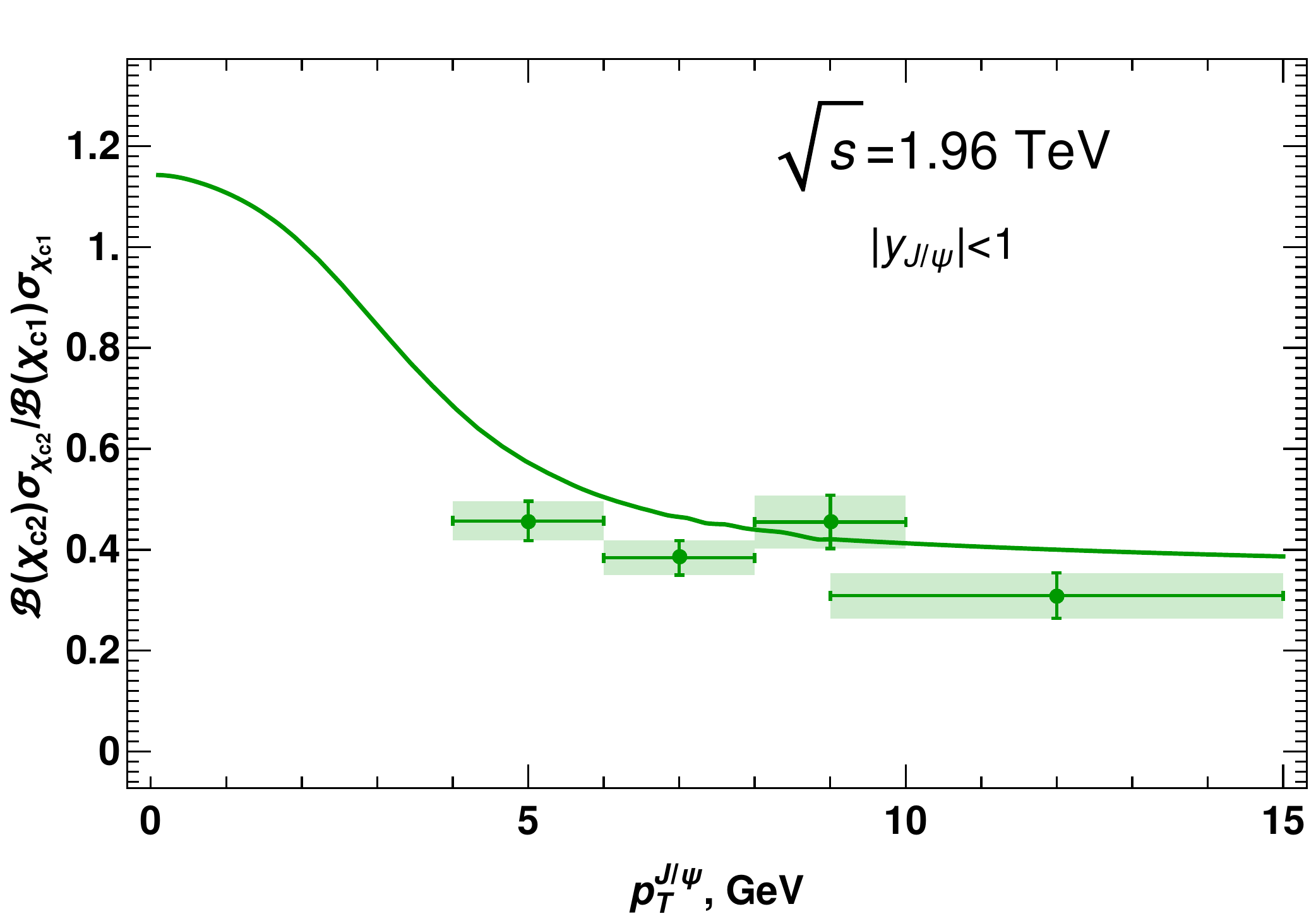}

\caption{\label{fig:ComparisonDataChic-CDF}Left: Comparison of the predicted
$p_{T}^{J/\psi}$-dependence for $\chi_{c}$ meson production
with experimental data from CDF~\cite{Abe:1997yz} at central rapidities
($|y|<1$). Right: Comparison of model predictions for the ratio of
the $\chi_{c2}$ and $\chi_{c1}$ cross-sections with experimental
data from~\cite{Abulencia:2007bra}.}
\end{figure}

In Figures~\ref{fig:ComparisonDataChic},~\ref{fig:ComparisonDataChic-CDF}
we compare the model predictions for the $\chi_{cJ}$ production with
available data from ATLAS~\cite{ATLAS:2014ala}, CMS~\cite{Chatrchyan:2012ub},
LHCb~\cite{Aaij:2013dja} and CDF~\cite{Abe:1997yz}. We can see
that the CGC/Sat model provides a reasonable description of the available
data in a wide kinematic range and thus might be used for further
analysis. 

\begin{figure}
\includegraphics[width=9cm]{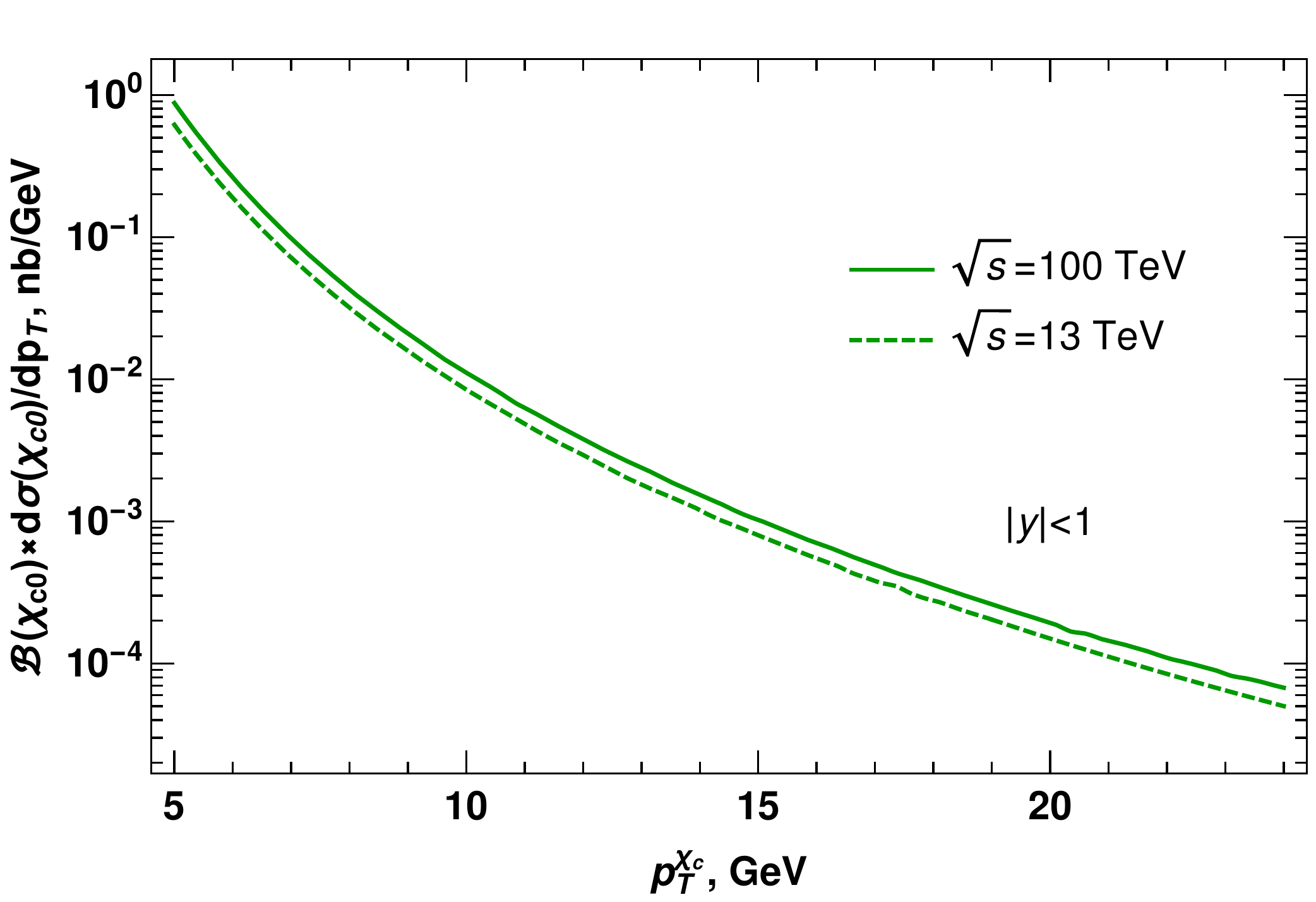}\includegraphics[width=9cm]{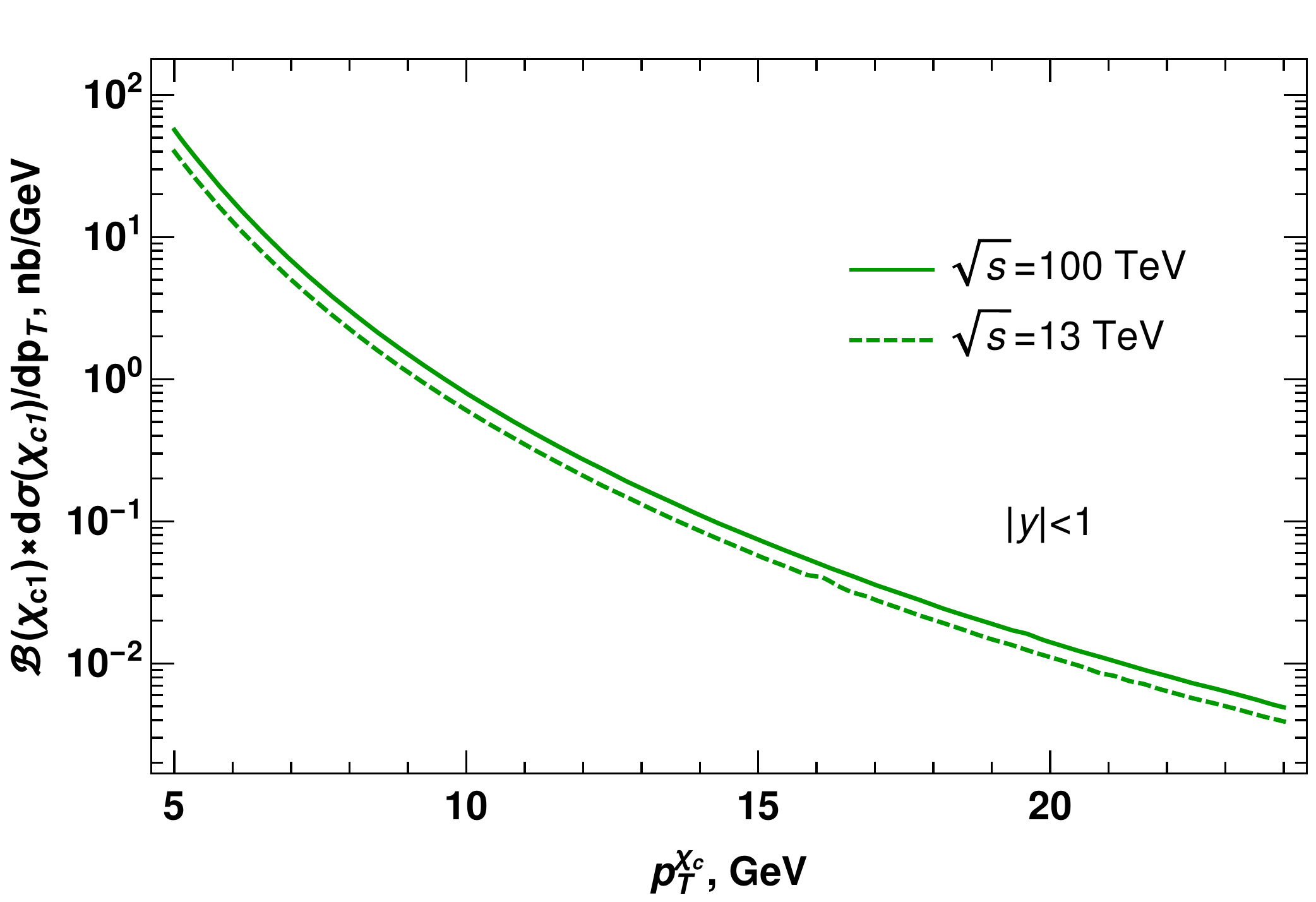}

\includegraphics[width=9cm]{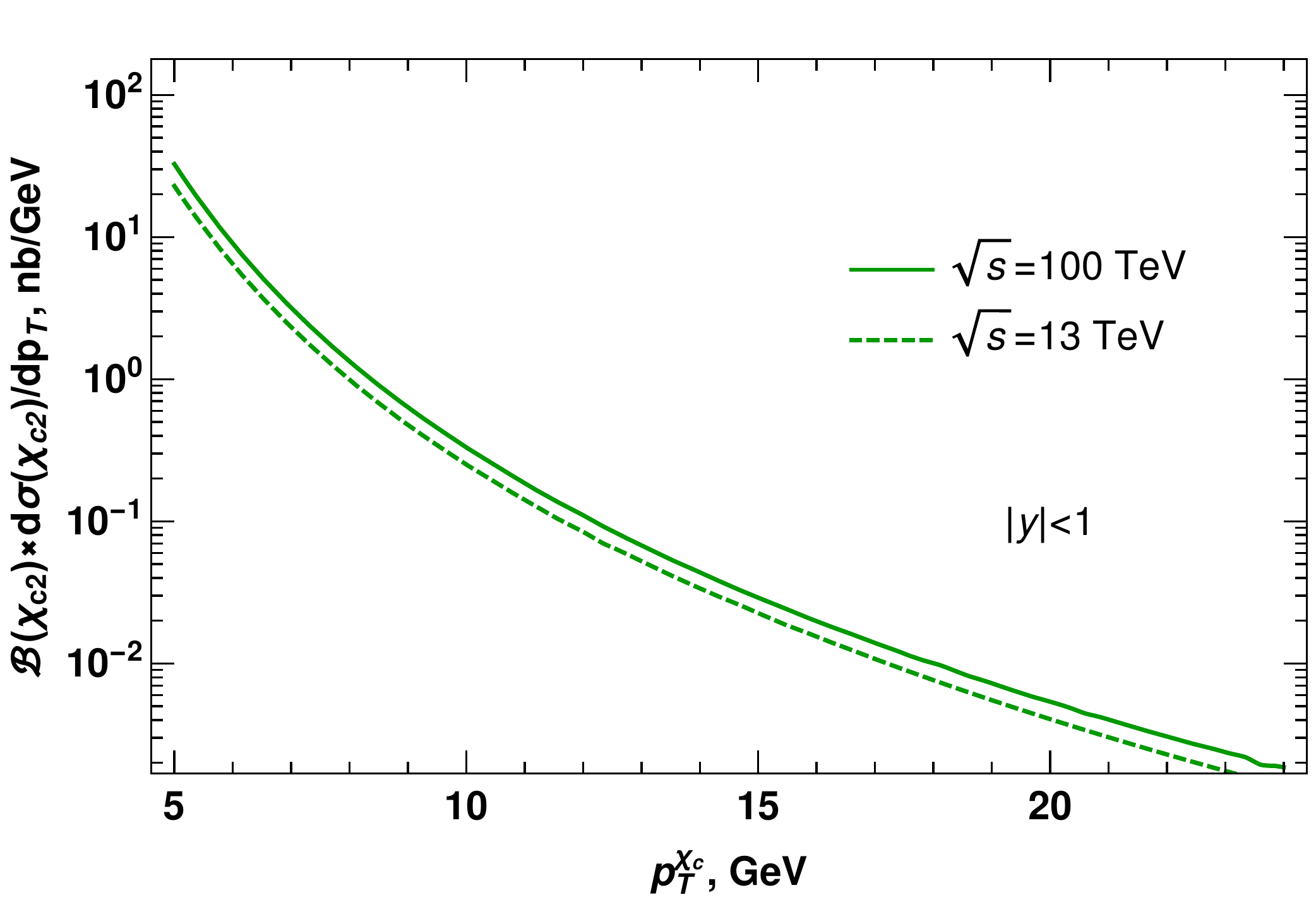}\includegraphics[width=9cm]{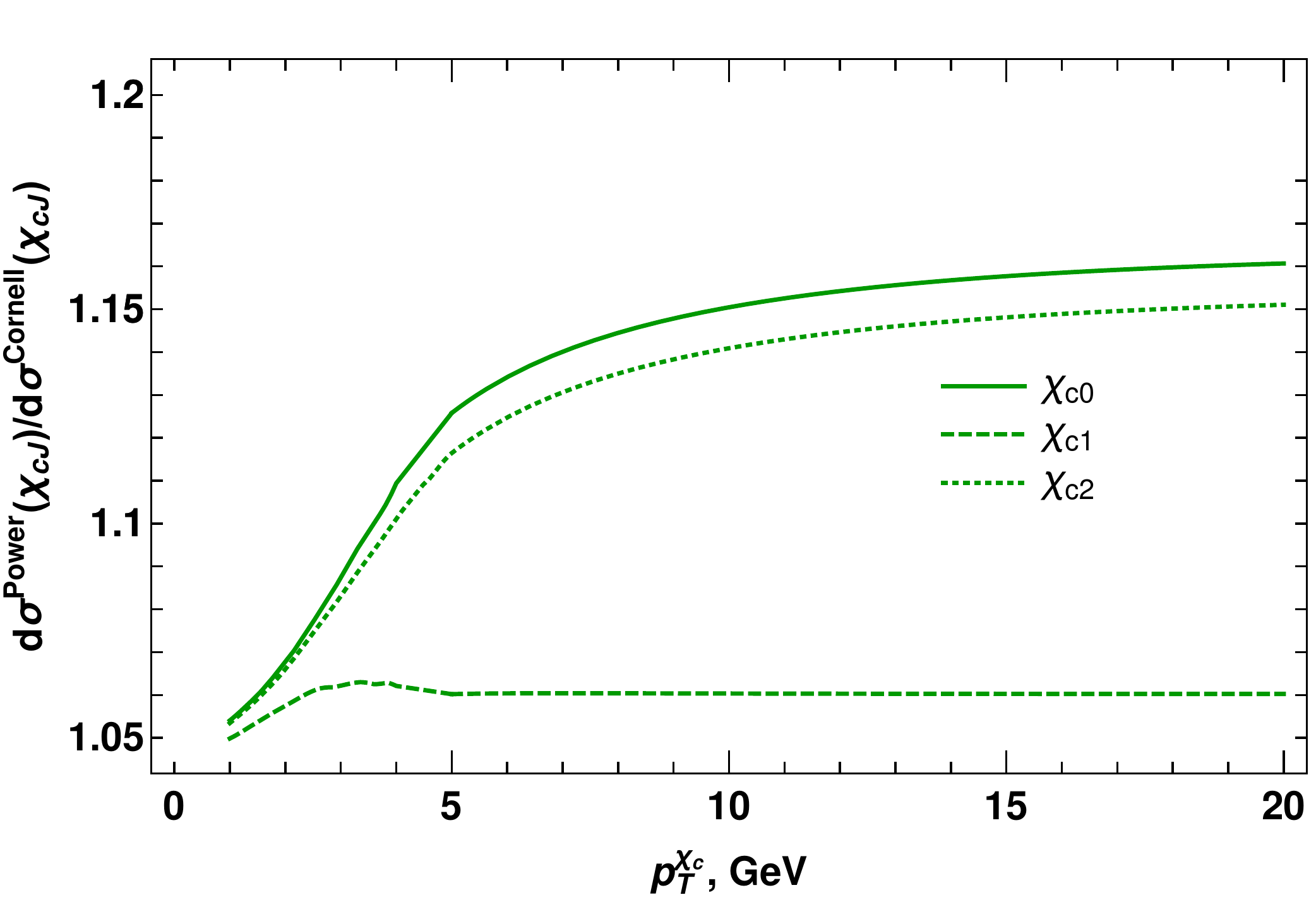}

\caption{\label{fig:sDependence}Upper row and left plot in lower row: Comparison
of the predicted $p_{T}$-dependence for $\chi_{c0}$, $\chi_{c1}$
and $\chi_{c2}$ cross-sections, for different values of $\sqrt{s}$.
All the theoretical curves are shown multiplied
by the branching $\mathcal{B}\left(\chi_{cJ}\right)\equiv Br\left(\chi_{cJ}\to\gamma+J/\psi\right)Br\left(J/\psi\to\mu^{+}\mu^{-}\right)$.
The cross-sections for $\chi_{c0}$ are strongly suppressed compared
to $\chi_{c1},\,\chi_{c2}$, due to differences in branching fractions
$Br\left(\chi_{cJ}\to\gamma+J/\psi\right)$. Right lower corner: The
ratio of cross-sections, evaluated with Cornell~\cite{Eichten:1978tg,Eichten:1979ms}
and Power-like~\cite{Martin:1980jx} parametrizations of the potential~(see
also Appendix~\ref{sec:WFs} for more details). }
\end{figure}

In Figure~\ref{fig:sDependence} we show the $p_{T}$-dependence
of the cross-sections at different values of the collision energies
$\sqrt{s}$, which might be relevant for the future experimental data.
We expect that the cross-section will proportionally increase as a
function of energy, without changing its shape. In order to illustrate
the dependence on the choice of wave function ($\sim$potential
model used for its evaluation), we have also shown in the same Figure
the ratio of the cross-sections evaluated with Cornell~\cite{Eichten:1978tg,Eichten:1979ms}
and Power-like~\cite{Martin:1980jx} parametrizations of the rest
frame potential. While the cross-sections changes several orders of
magnitude in the considered range of $p_{T}$, the uncertainty due
to choice of the potential does not exceed $15$ per cent. We also
got similar estimates for the parametrization~\cite{Quigg:1977dd}.
The cross-sections of $\chi_{c0}$ and $\chi_{c2}$ get large contributions
from the configuration with aligned spins of the quarks, whereas in the
case of $\chi_{c1}$ there is also a sizable contribution from configurations
where spins are anti-aligned, which explains the difference.

\begin{figure}
\includegraphics[width=9cm]{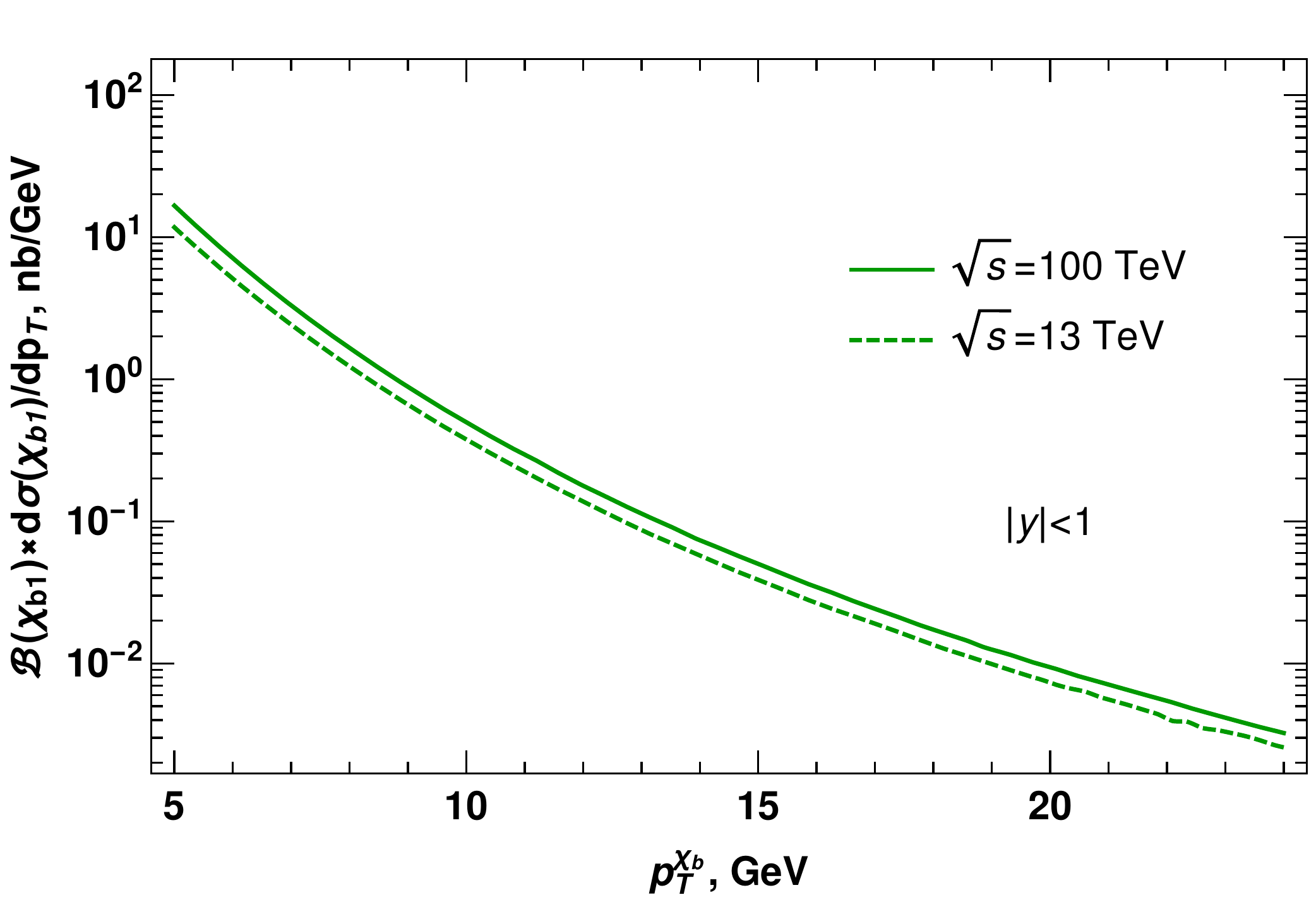}\includegraphics[width=9cm]{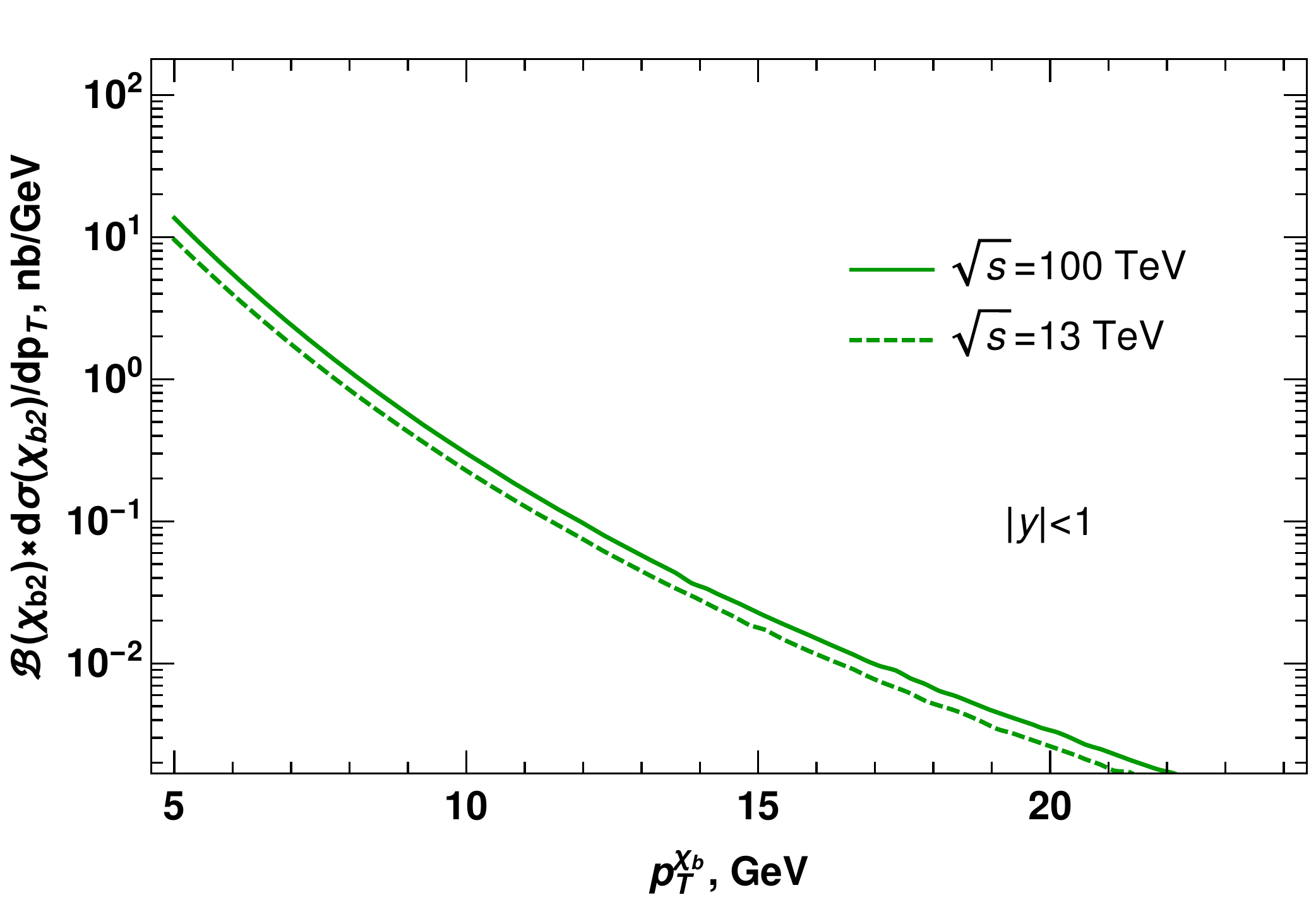}

\caption{\label{fig:Chib}The $p_{T}$-dependence for $\chi_{b1}$ and $\chi_{c2}$
cross-sections for different values of $\sqrt{s}$. All the theoretical
curves are shown multiplied by the branching $\mathcal{B}\left(\chi_{bJ}\right)\equiv Br\left(\chi_{bJ}\to\gamma+\Upsilon\right)Br\left(\Upsilon\to\mu^{+}\mu^{-}\right)$.}
\end{figure}

In Figure~\ref{fig:Chib} we show our predictions for the
$p_{T}$-dependence of $\chi_{b1}$ and $\chi_{b2}$ mesons. The estimated
cross-sections are smaller than for the $\chi_{c1}$ and $\chi_{c2}$
mesons, although within the reach of LHC experiments. So far there is no
published data from LHC for the cross-sections of these mesons, but
we hope that in the near future such measurements could be carried out.

\section{Multiplicity dependence}

\label{sec:Multiplicity}As we found in the previous section, the
CGC/Sat model~(\ref{FD1}) provides a reasonable description of the
$\chi_{c1}$ and $\chi_{c2}$ production data, at Tevatron and LHC kinematics.
The description of the multiplicity dependence presents more challenges
at the conceptual level, because there are different mechanisms to
produce enhanced number of charged particles $N_{{\rm ch}}$. The
probability of multiplicity fluctuations decreases rapidly as a function
of the number of produced charged particles $N_{{\rm ch}}$~\cite{Abelev:2012rz},
therefore for the study of the multiplicity dependence it is more
common to use a self-normalized ratio~\cite{Thakur:2018dmp} 
\begin{align}
\frac{dN_{M}/dy}{\langle dN_{M}/dy\rangle}\,\,=\frac{w\left(N_{M}\right)}{\left\langle w\left(N_{M}\right)\right\rangle }\,\frac{\left\langle w\left(N_{{\rm ch}}\right)\right\rangle }{w\left(N_{{\rm ch}}\right)}= & \frac{d\sigma_{M}\left(y,\,\eta,\,\sqrt{s},\,n\right)/dy}{d\sigma_{M}\left(y,\,\eta,\,\sqrt{s},\,\langle n\rangle=1\right)/dy}\bigg/\frac{d\sigma_{{\rm ch}}\left(\eta,\,\sqrt{s},\,Q^{2},\,n\right)/d\eta}{d\sigma_{{\rm ch}}\left(\eta,\,\sqrt{s},\,Q^{2},\,\langle n\rangle=1\right)/d\eta}\label{eq:NDef}
\end{align}
where $n=N_{{\rm ch}}/\langle N_{{\rm ch}}\rangle$ is the relative
enhancement of the charged particles in the bin, $w\left(N_{M}\right)/\left\langle w\left(N_{M}\right)\right\rangle $
and $w\left(N_{{\rm ch}}\right)/\left\langle w\left(N_{{\rm ch}}\right)\right\rangle $
are the self-normalized yields of quarkonium $M$ and charged particles
(minimal bias) events in a given multiplicity class; $d\sigma_{M}(y,\,\sqrt{s},\,n)$
is the production cross-sections for $M$, with rapidity $y$ and $\langle N_{{\rm ch}}\rangle=\Delta\eta\,dN_{{\rm ch}}/d\eta$
charged particles in the pseudorapidity window $(\eta-\Delta\eta/2,\,\,\eta+\Delta\eta/2)$.
If the inclusive cross-section of the process $pp\to M+X$ is proportional
to probability to produce a meson $M$ in a single $pp$ collision,
then the ratio~(\ref{eq:NDef}) gives a \emph{conditional} probability
to produce a meson $M$ in a $pp$ collision in which $N_{{\rm ch}}$
charged particles are produced. Due to the Local Parton-Hadron Duality
(LPHD) hypothesis~\cite{LPHD1,LPHD2,LPHD3}, the number of produced
charged particles is directly proportional to the number of partons
which stem from the individual pomerons and thus might be studied
using perturbative methods.

In the color dipole approach analyzed in this paper, we expect that
the multiplicity dependence is enhanced due to a large average number
of particles produced from each pomeron. Nevertheless, we still expect
that each such cascade (``pomeron'') should satisfy the nonlinear
Balitsky-Kovchegov equation, and therefore we expect that the dipole
amplitude~(\ref{eq:CGCDipoleParametrization}) should maintain its
form, although the value of the saturation scale $Q_{s}$ might be modified.
As was demonstrated in~\cite{KOLEB,KLN,DKLN}, the observed number
of charged multiplicity $dN_{{\rm ch}}/dy$ of soft hadrons in $pp$
collisions is given by the so-called KLN-style formula
\begin{equation}
\frac{dN_{{\rm ch}}}{dy}\,\,=\,\,c\,N_{I\!\!P}\,\frac{Q_{s}^{2}}{\bar{\alpha}_{S}\left(Q_{s}^{2}\right)}\label{MULTQS-2}
\end{equation}
where $c$ is a numerical coefficient, and $N_{I\!\!P}$ is the number
of BK pomerons. Solving (\ref{MULTQS-2}) algebraically, we could
extract $Q_{s}^{2}$ as a function of $dN_{{\rm ch}}/dy$. Taking
into account that the distribution $dN_{{\rm ch}}/dy$ is almost flat,
we may approximate $n=N_{{\rm ch}}/\langle N_{{\rm ch}}\rangle\approx(dN_{{\rm ch}}/dy)/\langle dN_{{\rm ch}}/dy\rangle$,
so (\ref{MULTQS-2}) allows to express $Q_{s}^{2}$ as a function
of $n$. Usually in the literature the logarithmic dependence on
$n$, which stems from the running coupling in the denominator of~(\ref{MULTQS-2})
is disregarded, so (\ref{MULTQS-2}) reduces to a simpler linearly
growing dependence on $n$~\cite{KOLEB,KLN,DKLN,Kharzeev:2000ph,Kovchegov:2000hz,LERE,Lappi:2011gu},

\begin{equation}
Q_{s}^{2}\left(x,\,b;\,n\right)\,\,=\,\,n\,Q^{2}\left(x,\,b\right).\label{QSN-1}
\end{equation}
The precision of this assumption was tested in~\cite{Ma:2018bax},
and it was found that a numerical solution of the running coupling
Balitsky-Kovchegov (rcBK) equation differs from the approximate~(\ref{QSN-1})
by less than 10\% in the region of interest ($n\lesssim10$). This
correction is within the precision of current evaluations, and for this
reason in what follows we will use~(\ref{QSN-1}) for our estimates.
While at LHC energies it is expected that the typical values of saturation
scale $Q_{s}\left(x,\,b\right)$ fall into the range 0.5-1 ${\rm GeV}$,
from~(\ref{QSN-1}) we can see that in events with enhanced multiplicity
this parameter might exceed the values of heavy quark mass $m_{Q}$
and lead to an interplay of large-$Q_{s}$ and large-$m_{Q}$ limits.
Since increase of multiplicity and increase of energy (decrease of
$x$) affect $Q_{s}^{2}$ in a similar way, the study of the high-multiplicity
events allows to study a deeply saturated regime which determines
the dyanamics of all processes at significantly higher energies. 

For phenomenological estimates of the multiplicity dependence it is
very important that the rapidity bins used for collecting the co-produced
charged hadrons overlaps with the rapidity bins used for collection of
quarkonia. As was illustrated in our previous publications~\cite{LESI,Schmidt:2020fgn,Siddikov:2019xvf},
the observed enhanced multiplicity should be shared equally between
all cut pomerons which \emph{could} contribute to co-production of
hadrons in a given experimental setup. For this reason the strongest
multiplicity dependence will show up in the case when both quarkonia
and hadrons are collected at central rapidities. In what follows we
will focus on this setup. 

\begin{figure}
\includegraphics[width=9cm]{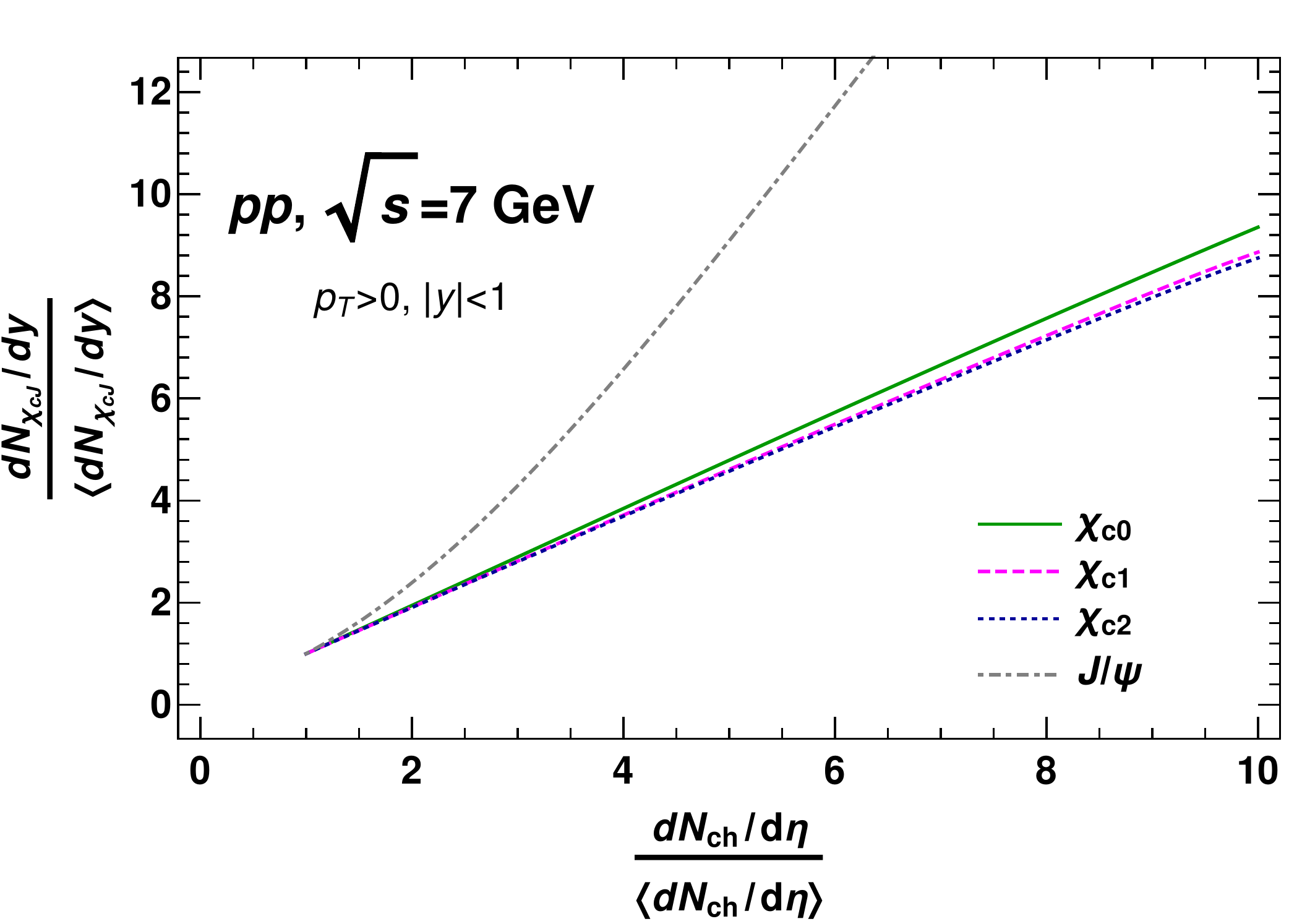}\includegraphics[width=9cm]{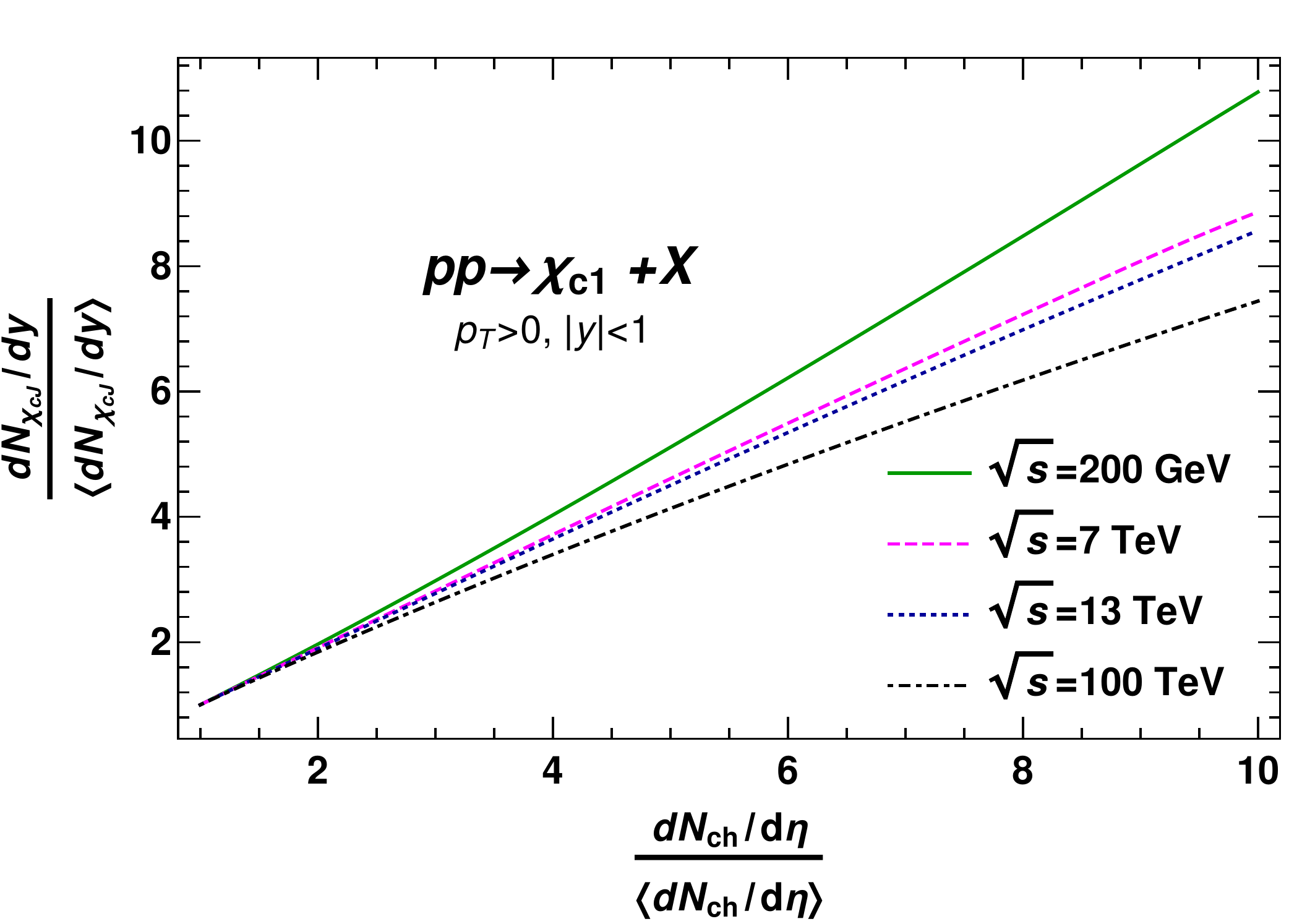}

\caption{\label{fig:Multiplicity-Chic}Left: multiplicity dependence for different
$\chi_{cJ}$ states. While the cross-sections differ quite significantly
due to spin structure, the self-normalized ratios are very close to
each other. For the sake of reference we also added a dot-dashed grey
curve for $J/\psi$ production from our previous~\cite{LESI,Siddikov:2019xvf}.
Right: Dependence of the multiplicity shapes on collision energies
$\sqrt{s}$. The plot is done for $\chi_{c1}$ meson production, but the
results for $\chi_{c0},\,\chi_{c2}$ are almost identical. All evaluations
are done assuming that charged particles and quarkonia are collected
at central rapidities ($|\eta,y|<1$), similar to what is available
for $J/\psi$ production from~\cite{Khatun:2019slm}.}
\label{FwdCentral-1} 
\end{figure}

\begin{figure}
\includegraphics[width=9cm]{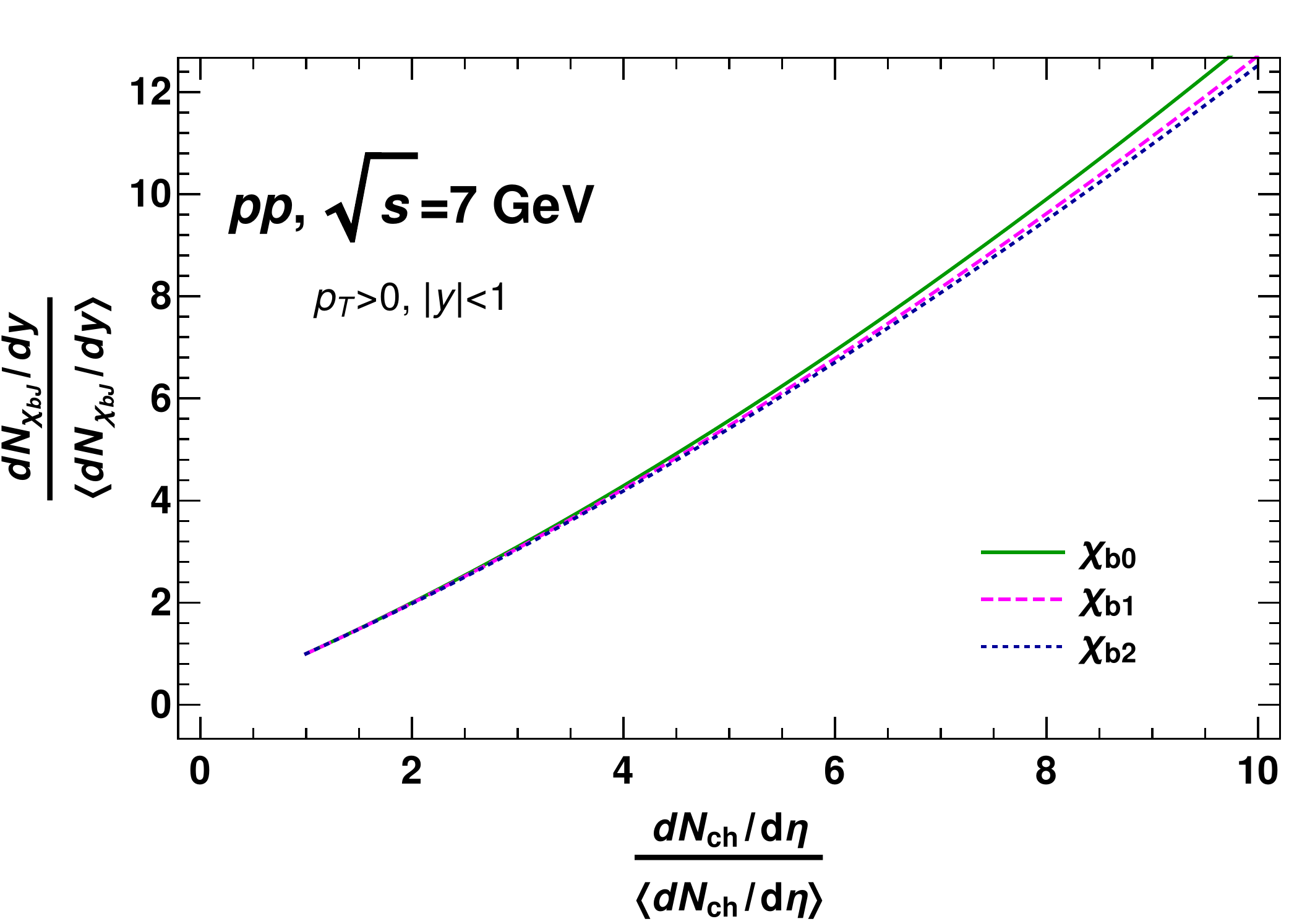}\includegraphics[width=9cm]{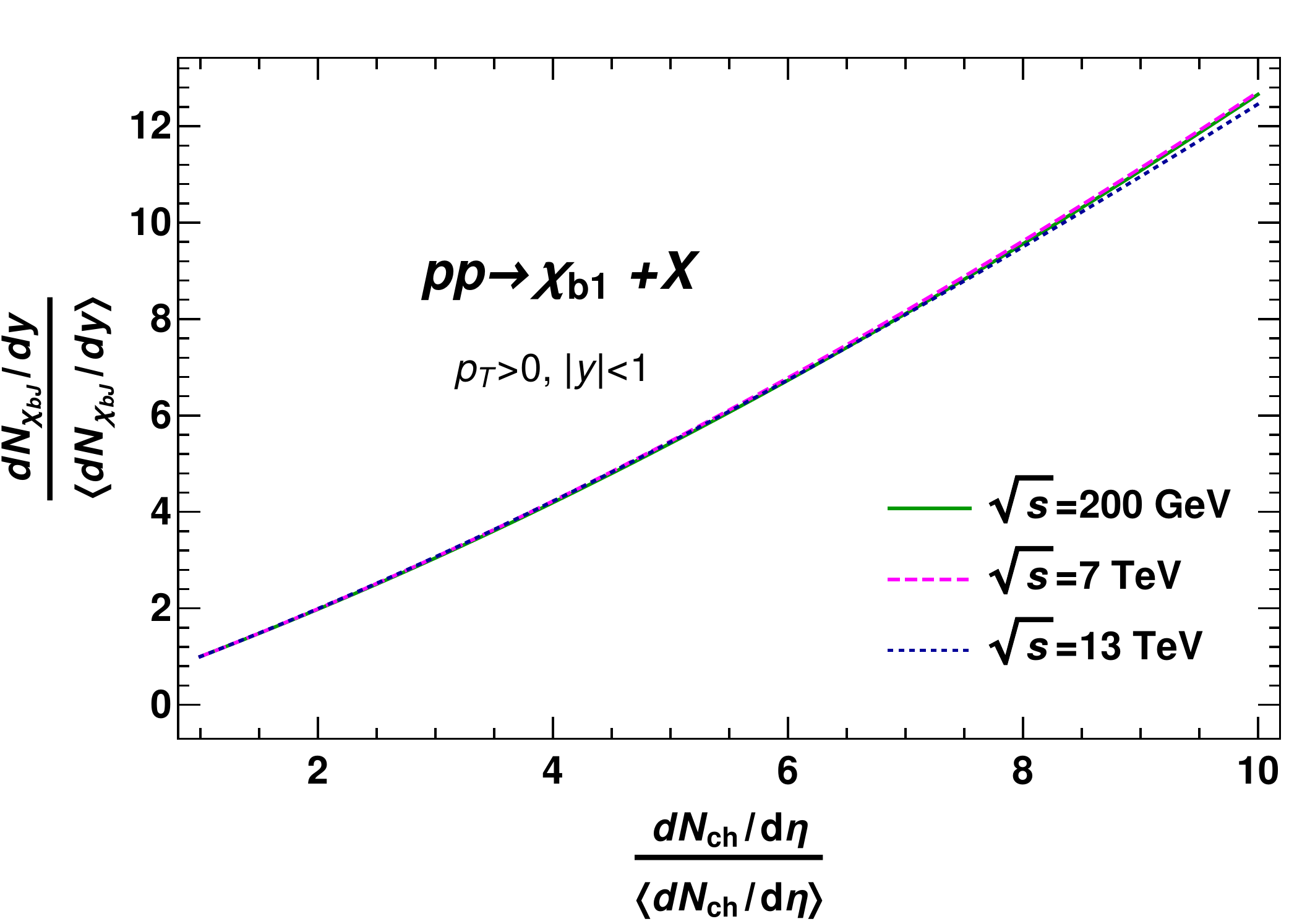}

\caption{\label{fig:Multiplicity-Chib}Left: multiplicity dependence for different
$\chi_{bJ}$ states. While the cross-sections differ quite significantly
due to spin structure, the self-normalized ratios are very close to
each other. Right: Dependence of the multiplicity shapes on collision
energies $\sqrt{s}$. The plot is done for $\chi_{b1}$ meson production,
but results for $\chi_{b0},\,\chi_{b2}$ are almost identical. All
evaluations are done assuming that charged particles and quarkonia
are collected at central rapidities ($|\eta,y|<1$), similar to what
is available from~\cite{Khatun:2019slm}. }
\end{figure}

In Figures~\ref{fig:Multiplicity-Chic},~\ref{fig:Multiplicity-Chib}
we show the multiplicity dependence of the $\chi_{cJ}$ and
$\chi_{bJ}$ mesons for different energies. As we can see, the dependence
is much milder than that of the $1S$ quarkonia (dot-dashed curve
with label ``$J/\psi$''), in agreement with our expectations based
on dominance of two-pomeron mechanism. From~(\ref{QSN-1}) and the
structure of the dipole cross-section~(\ref{eq:CGCDipoleParametrization})
we can expect that each cut pomeron contributes to the multiplicity dependence
factor $\sim n^{\langle\gamma_{{\rm eff}}\rangle}$, where the parameter
$\gamma_{{\rm eff}}$ was defined in~(\ref{eq:gamma_eff}). Since
$\chi_{b}$ has smaller size than $\chi_{c}$, the typical values
of $\langle\gamma_{{\rm eff}}\rangle$ are larger for the former than
for the latter, and $\chi_{b}$ has slightly faster dependence
on multiplicity than $\chi_{c}$. Similarly, we can understand the
change of multiplicity dependence with energy: due to prefactor $1/Y$
in~(\ref{eq:gamma_eff}), the average values of the parameter $\langle\gamma_{{\rm eff}}\rangle$
\emph{decrease} as a function of $\sqrt{s}$, and for this reason the dependence on multiplicity becomes
milder for larger energies $\sqrt{s_{pp}}$.

\section{Conclusions}

\label{sec:Concusion} In this paper we analyzed in detail the production
of $\chi_{c}$ and $\chi_{b}$ mesons in the CGC/Sat approach. We
found that the model predictions for the $p_{T}$-dependent cross-section
are in agreement with available experimental data for $\chi_{c1}$
and $\chi_{c2}$ mesons in LHC kinematics. We also made predictions
for $\chi_{b}$ mesons, which might be checked in the ongoing and future
experiments, both at RHIC and at LHC. We also studied the dependence
of the cross-sections on multiplicity of co-produced hadrons, and
found that it is significantly milder than that of $1S$ quarkonia
($J/\psi,\,\Upsilon...$). This happens because the dominant production
mechanism of for the $P$-wave quarkonia is the two-pomeron fusion,
whereas the three-pomeron contributions are strongly suppressed at
high energies. Our evaluation is largely parameter-free and relies
only on the choice of the parametrization for the dipole cross-section~(\ref{eq:CGCDipoleParametrization})
and the wave function of the meson. 

The explanation of multiplicity dependence in the CGC/Sat approach differs
from other  approaches suggested for the description of multiplicity dependence,
like \emph{e.g}. the percolation approach~\cite{PER} or modification
of the slope of the elastic amplitude~\cite{Kopeliovich:2013yfa}.
While for $1S$ quarkonia all approaches give comparable descriptions,
this is not so for $P$-wave quarkonia. For this reason we expect
that the measurement of the multiplicity dependence of $\chi_{c}$
and $\chi_{b}$ would be an important litmus test for all the models
which describe production of quarkonia, and we hope that it will be done
both at LHC and RHIC.

\section{Acknowledgements}

We thank our colleagues at UTFSM university for encouraging discussions.
This research was partially supported by the project ANID PIA/APOYO
AFB180002\emph{ }(Chile) and Fondecyt (Chile) grant 1180232. Also,
we thank Yuri Ivanov for technical support of the USM HPC cluster,
where some evaluations were performed.

\appendix

\section{Evaluation of the dipole amplitudes}

\label{sec:Derivation} 
\begin{figure}
\includegraphics[width=9cm]{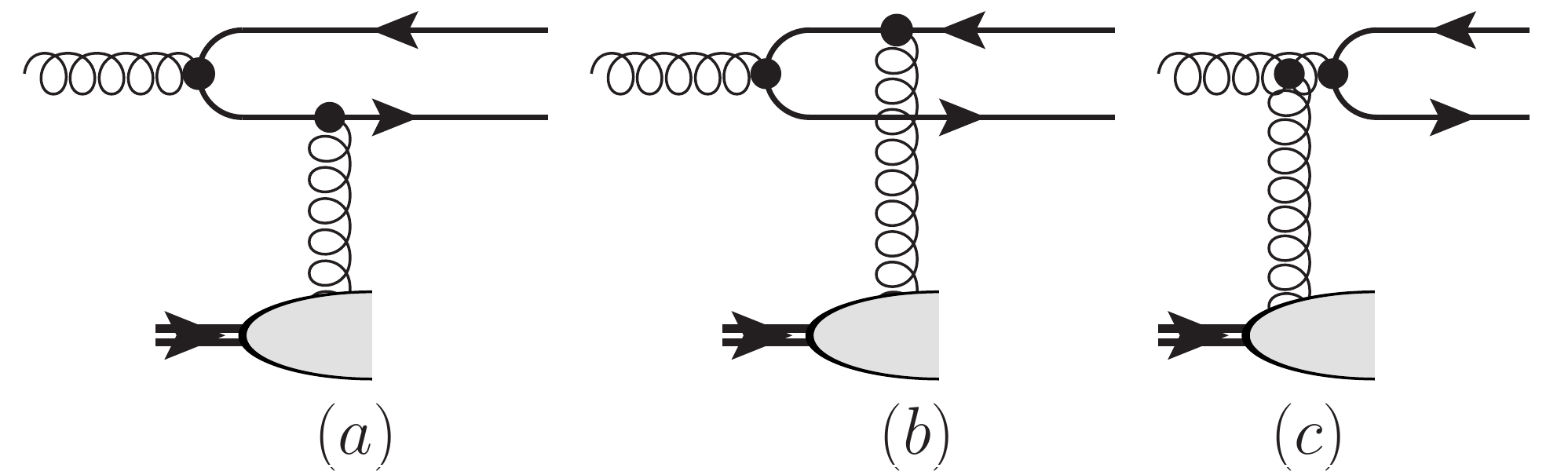}

\caption{\label{fig:Dipole2Pomeron}The diagrams which contribute to the heavy
meson production cross-section in the leading order perturbative QCD.
The contribution of the last diagram ($c$) to the meson formation
might be also viewed as gluon-gluon fusion $gg\to g$, with subsequent
gluon fragmentation $g\to\bar{Q}Q$. In CGC parametrization of the
dipole cross-section approach each ``gluon'' is replaced with reggeized
gluon (BK pomeron), which satisfies the Balitsky-Kovchegov equation
and corresponds to a fan-like shower of soft particles.}
\end{figure}

In this Appendix, for the sake of completeness, we explain the main
technical steps and assumptions used for the derivation of the cross-section~(\ref{FD1}).
The general rules, which allow to express the cross-sections of hard
processes in terms of the color singlet dipole cross-section, might
be found in~\cite{GLR,McLerran:1993ni,McLerran:1993ka,McLerran:1994vd,MUQI,MV,gbw01:1,Kopeliovich:2002yv,Kopeliovich:2001ee}.
In the heavy quark mass limit the strong coupling $\alpha_{s}(m_{Q})$
is small, so the interaction of a heavy $\bar{Q}Q$ dipole with gluons
might be considered perturbatively. At the same time, we tacitly assume
that each such gluon should be understood as a parton shower (``pomeron'').

In the high-energy eikonal picture, the interaction of the quarks
and antiquark with a $t$-channel gluon are described by a factor
$\pm ig\,t^{a}\gamma\left(\boldsymbol{x}_{\perp}\right)$, where $\boldsymbol{x}_{\perp}$
is the transverse coordinate of the quark, and the function $\gamma\left(\boldsymbol{x}_{\perp}\right)$
is related to a distribution of gluons in the target. This function
is related to a dipole scattering amplitude $N(x,\,\boldsymbol{r})$
as 
\begin{equation}
\Delta N(x,\,\boldsymbol{r})\equiv N(x,\,\infty)-N(x,\,\boldsymbol{r})=\frac{1}{8}\int d^{2}b\left|\gamma\left(x,\,\boldsymbol{b}-z\boldsymbol{r}\right)-\gamma\left(x,\,\boldsymbol{b}+\bar{z}\boldsymbol{r}\right)\right|^{2}\label{eq:DipoleX}
\end{equation}
where $\boldsymbol{r}$ is the transverse size of the dipole, and
$z$ is the light-cone fraction of the dipole momentum carried by
the quarks. The equation~(\ref{eq:DipoleX}) might be rewritten in
the form 
\begin{equation}
\frac{1}{8}\int d^{2}\boldsymbol{b}\gamma(x,\,\boldsymbol{b})\gamma(x,\,\boldsymbol{b}+\boldsymbol{r})=\frac{1}{2}N(x,\,\boldsymbol{r})+\underbrace{\int d^{2}b\,\left|\gamma(x,\,\boldsymbol{b})\right|^{2}-\frac{1}{2}N(x,\,\infty)}_{={\rm const}}.\label{eq:SigmaDef}
\end{equation}
The value of the constant is related to the infrared behavior of the
theory, and for the observables which we consider in this paper, it
cancels exactly. For very small dipoles, the dipole scattering amplitude
is related to the gluon uPDF as~\footnote{In the literature definitions of the unintegrated PDF $\mathcal{F}\left(x,\,k_{\perp}\right)$
might differ by a factor $k_{\perp}^{2}$.} 
\begin{equation}
N\left(x,\,\vec{\boldsymbol{r}}\right)=\frac{4\pi\alpha_{s}}{3}\int\frac{d^{2}k_{\perp}}{k_{\perp}^{2}}\mathcal{F}\left(x,\,k_{\perp}\right)\left(1-e^{ik\cdot r}\right)+\mathcal{O}\left(\frac{\Lambda_{{\rm QCD}}}{m_{c}}\right),\label{eq:Dip}
\end{equation}
so the functions $\gamma\left(x,\,\boldsymbol{r}\right)$ might be
also related to the unintegrated gluon densities in coordinate space.
With the help of~(\ref{eq:SigmaDef}), it is possible to express
the exclusive amplitudes and inclusive cross-sections as linear combinations
of the \emph{color singlet} dipole cross-sections $\sigma(x,\,\boldsymbol{r})$
with different arguments. While in the deeply saturated regime we
can no longer speak about individual gluons (or pomerons), we expect
that the relations between the dipole amplitudes and color singlet
cross-sections should be valid even in this case.

For the case of $P$-wave production, we should take into account that
the quark and antiquark transverse coordinates are given by $\boldsymbol{b}_{i}-z_{i}\,\boldsymbol{r}_{i}$
and $\boldsymbol{b}_{i}+\bar{z}_{i}\,\boldsymbol{r}_{i}$ respectively,
where subindex $i$ takes values $i=1,\,2$ for the amplitude and
its conjugate. For the evaluation of the $p_{T}$-dependent cross-section
we need to project the coordinate space quark distribution onto the
state with definite transverse momentum $\boldsymbol{p}_{T}$, so
we have to evaluate the additional convolution $\sim\int d^{2}\boldsymbol{b}_{1}d^{2}\boldsymbol{b}_{2}\,e^{i\boldsymbol{p}_{T}\cdot\left(\boldsymbol{b}_{1}-\boldsymbol{b}_{2}\right)}\equiv\int d^{2}\boldsymbol{b}_{21}d^{2}\boldsymbol{b}\,e^{-i\boldsymbol{p}_{T}\cdot\boldsymbol{b}_{21}}$,
where $\boldsymbol{b}_{21}\equiv\boldsymbol{b}_{2}-\boldsymbol{b}_{1}$
is the difference of impact parameters of the dipole center of mass
in the amplitude and its conjugate, and $\boldsymbol{b}=(\boldsymbol{b}_{1}+\boldsymbol{b}_{2})/2$.
The integral over $\int d^{2}\boldsymbol{b}$ is evaluated using~(\ref{eq:SigmaDef}),
and after simple algebraic simplifications we can get~(\ref{FD1}).
For the three-pomeron we consider that in a target rest frame we have
a scattering of a dipole in the field of two pomerons. Since at high
energies each pomeron reggeizes independently, we expect that the amplitude
will be a mere product of the scattering amplitudes on each pomeron,
thus yielding~(\ref{eq:N3Def}).

All evaluations of this appendix were done in the frame where the
momentum of the primordial gluon is zero. In any other frame we should
take into account an additional convolution with the momentum distribution
of the incident (``primordial'') gluons, as appears in the first
line of~(\ref{FD1}).

\section{Wave functions}

\label{sec:WFs}

For evaluations of~\ref{FD1}, we will need explicit parametrizations
for the $\bar{Q}Q$ component of the light cone gluon wave function
$\Psi_{\bar{Q}Q}$ and the $P$-wave quarkonia wave function $\Psi_{M}$.
We may expect that in the heavy quark mass limit for $\Psi_{\bar{Q}Q}$
we may use the well-known perturbative expressions available from
the literature~\cite{Dosch:1996ss,Bjorken:1970ah}, 
\begin{align}
\Psi_{\bar{Q}Q}^{(+1)}(z,\,\boldsymbol{r}) & =\frac{\sqrt{2}}{2\pi}\left(ie^{i\theta}\varepsilon\left(z\,\delta_{h+}\delta_{\bar{h}-}-(1-z)\delta_{h-}\delta_{\bar{h}+}\right)K_{1}(\varepsilon r)+m_{Q}\delta_{h+}\delta_{\bar{h}+}K_{0}(\varepsilon r)\right),\label{eq:WFDef}\\
\Psi_{\bar{Q}Q}^{(-1)}(z,\,\boldsymbol{r}) & =\frac{\sqrt{2}}{2\pi}\left(ie^{-i\theta}\varepsilon\left((1-z)\,\delta_{h+}\delta_{\bar{h}-}-z\delta_{h-}\delta_{\bar{h}+}\right)K_{1}(\varepsilon r)+m_{Q}\delta_{h-}\delta_{\bar{h}-}K_{0}(\varepsilon r)\right),\nonumber 
\end{align}
where the superscript index $(\pm1)$ of $\Psi_{\bar{Q}Q}$ refers
to helicity of the projectile gluon, and $h,\,\bar{h}$ in the r.h.s.
are the helicities of the quark and antiquark respectively. 

Most previous evaluations of quarkonia production~\cite{Likhoded:2014kfa,Baranov:2015yea,Jia:2014jfa,Hagler:2000dd,Babiarz:2020jkh,Cisek:2017gno,Diakonov:2012vb}
were done in the heavy quark mass limit, assuming that the wave function
$\Psi_{M}(z,\,\boldsymbol{r})$ might be approximated with its small-$\boldsymbol{r}$
Taylor expansion, which starts with a linear term for $P$-wave. In
this case the dependence on the wave function reduces to dependence
on the value of the slope $|\mathcal{R}'(0)|$. This scheme is justified
in the heavy quark mass limit. However, it is clear that for charmonia
the heavy quark mass limit might not work very well, and for this
reason we will maintain the full $r$-dependence of the wave function. 

For our evaluations we construct a light-cone wave function from the
rest frame wave functions evaluated in the potential models using
the Brodsky-Huang-Lepage (BHL) prescription~\cite{BHT} (see also
a similar scheme~\cite{Terentev:1976jk}). It is known that for heavy
quarkonia the results of the BHL prescription are close to the wave
functions evaluated in Covariant Spectator Theory~\cite{Stadler:2018hjv}
as well as lattice evaluations~\cite{Daniel:1990ah,Kawanai:2011xb,Kawanai:2013aca}.
According to BHL prescription, the light-cone wave function related
to the rest frame wave function as~\cite{BHT,Nemchik:1996cw,Hufner:2000jb}
\begin{align}
\Phi_{{\rm LC}}(z,\,r) & =\int d^{2}k_{\perp}e^{ik_{\perp}\cdot r_{\perp}}\left(\frac{k_{\perp}^{2}+m_{Q}^{2}}{2\,z^{3}(1-z)^{3}}\right)\psi_{{\rm RF}}\left(\sqrt{\frac{k_{\perp}^{2}+(1-2z)^{2}m_{Q}^{2}}{4z(1-z)}}\right),
\end{align}
where $\psi_{{\rm RF}}(\boldsymbol{k})$ is the Fourier image of the
rest frame wave function. Due to spin-orbital interaction, we expect
that the spinorial structure of the wave functions of $\chi_{cJ},\,\chi_{bJ}$
will be depend crucially on the angular momentum $J$, and for this
reason the production cross-sections of $\chi_{cJ},\chi_{bJ}$ mesons
will acquire dependence on $J$~\footnote{Technically, we expect that the spin-angular dependence will contribute
via overlap with helicity-dependent wave function of gluon~\ref{eq:WFDef}.}. 

The rest frame quarkonium wave function might be written using the
Clebsch-Gordan coefficients as 
\[
\bigg\langle\boldsymbol{r}\left|J,\,J_{z}\right\rangle =\sum_{S=0}^{1}\sum_{M+S_{z}=J_{z}}\sum_{s_{1}+s_{2}=S_{z}}\chi_{s_{1}}\bar{\chi}_{s_{2}}\left\langle S,S_{z}|\frac{1}{2},\,s_{1};\,\frac{1}{2},\,s_{2}\right\rangle \left\langle J,J_{z}|1,\,M;\,S,\,S_{z}\right\rangle \mathcal{R}_{n1}(r)Y_{1M}(\hat{\boldsymbol{r}}),
\]
where $\{\chi_{s}\}$ are spinors corresponding to definite projection
of spin $s$ of the quark and antiquark, $M$ is the projection of
orbital angular momentum, $\mathcal{R}_{n1}(r)$ is the radial wave
function of the $P$-wave quarkonium, and $Y_{1M}$ is an ordinary
spherical harmonic. A set of useful relations between Clebsch-Gordan
coefficients, which facilitate the summations, might be found in~\cite{Baranov:2015yea,Boer:2012bt}.
For the evaluations of radial part $\mathcal{R}_{n1}$ we used expressions
found in potential models. For the sake of comparison in evaluation
we considered three different choices of the potential:
\begin{itemize}
\item The Cornell potential which was introduced in~\cite{Eichten:1978tg,Eichten:1979ms},
\begin{equation}
V_{{\rm Cornell}}(r)=-\frac{\alpha}{r}+\sigma\,r.\label{eq:Cornell}
\end{equation}
\item The potential with logarithmic large-$r$ behaviour suggested in~\cite{Quigg:1977dd},
\begin{equation}
V_{{\rm Log}}(r)=\alpha+\beta\,\ln r.\label{eq:Log}
\end{equation}
\item The power-like potential introduced in~\cite{Martin:1980jx}, 
\begin{equation}
V_{{\rm pow}}(r)=a+b\,r^{\alpha}.\label{Power}
\end{equation}
\end{itemize}
We checked that the wave functions obtained with all three potentials
have similar shapes. As we mentioned earlier, in the heavy quark mass
limit we expect that the wave function might be approximated by its
behavior near the $r\approx0$. The wave functions of the $P$-wave
quarkonia have a node at $r\approx0$, so the relevant parameter, which
determines the cross-sections in this limit, is the value of the slope
$\left|\mathcal{R}'(0)\right|^{2}$. To facilitate comparison with
other approaches, in Table~\ref{tab:ValuesR} we provide the values
of the latter parameter.
\begin{table}
\begin{tabular}{|c|c|c|c|}
\hline 
 & Cornell~(\ref{eq:Cornell}) & Log~(\ref{eq:Log}) & Power~(\ref{Power})\tabularnewline
\hline 
$\left|\mathcal{R}'(0)\right|^{2},\,\left[{\rm GeV}^{5}\right]$ & 0.34 & 0.26 & 0.33\tabularnewline
\hline 
\end{tabular}\caption{\label{tab:ValuesR}Values of the slope $\left|\mathcal{R}'(0)\right|^{2}$
evaluated with different models of interquark potential interaction.}
\end{table}
For the evaluation of the overlap with gluon wave function~(\ref{eq:WFDef}),
it is convenient to rewrite the quarkonium wave functions in a helicity
basis. Conventionally, this is done  applying the Melosh-Wigner spin
rotation operators defined in~\cite{MELOSH1,MELOSH2}. 

 \end{document}